\documentclass{article}

\PassOptionsToPackage{numbers, compress}{natbib}
\usepackage[main, final]{neurips_2026}

\usepackage[utf8]{inputenc}
\usepackage[T1]{fontenc}
\usepackage{hyperref}
\usepackage{url}
\usepackage{booktabs}
\usepackage{amsfonts}
\usepackage{nicefrac}
\usepackage{microtype}
\usepackage{xcolor}
\usepackage{xspace}
\usepackage{float}
\usepackage{amsmath}

\usepackage{graphicx}
\usepackage{multirow}
\usepackage{placeins}

\newcommand{\ours}{StereoSplat\xspace}
\newcommand{\oursdataset}{XRStereo\xspace}

\newcommand{\boldparagraph}[1]{\vspace{2pt}\noindent{\bf #1}}

\title{StereoSplat: Metric-Scale Novel View Synthesis via Stereo-Grounded Gaussian Splatting}

\author{%
  Vladimir Yugay$^{1}$\thanks{Work done during an internship at Meta.} \quad
  Denis Rozumny$^{2}$ \quad
  Theo Gevers$^{1}$ \\
  \bfseries Elias Vansteenkiste$^{2}$ \quad
  Martin R. Oswald$^{1}$ \quad
  Diogo C. Luvizon$^{2}$ \\
  $^{1}$University of Amsterdam \qquad $^{2}$Meta \\
  \url{https://vladimiryugay.github.io/stereosplat}
}

\begin{document}
\maketitle
\begin{figure}[H]
  \centering
  \includegraphics[width=\linewidth]{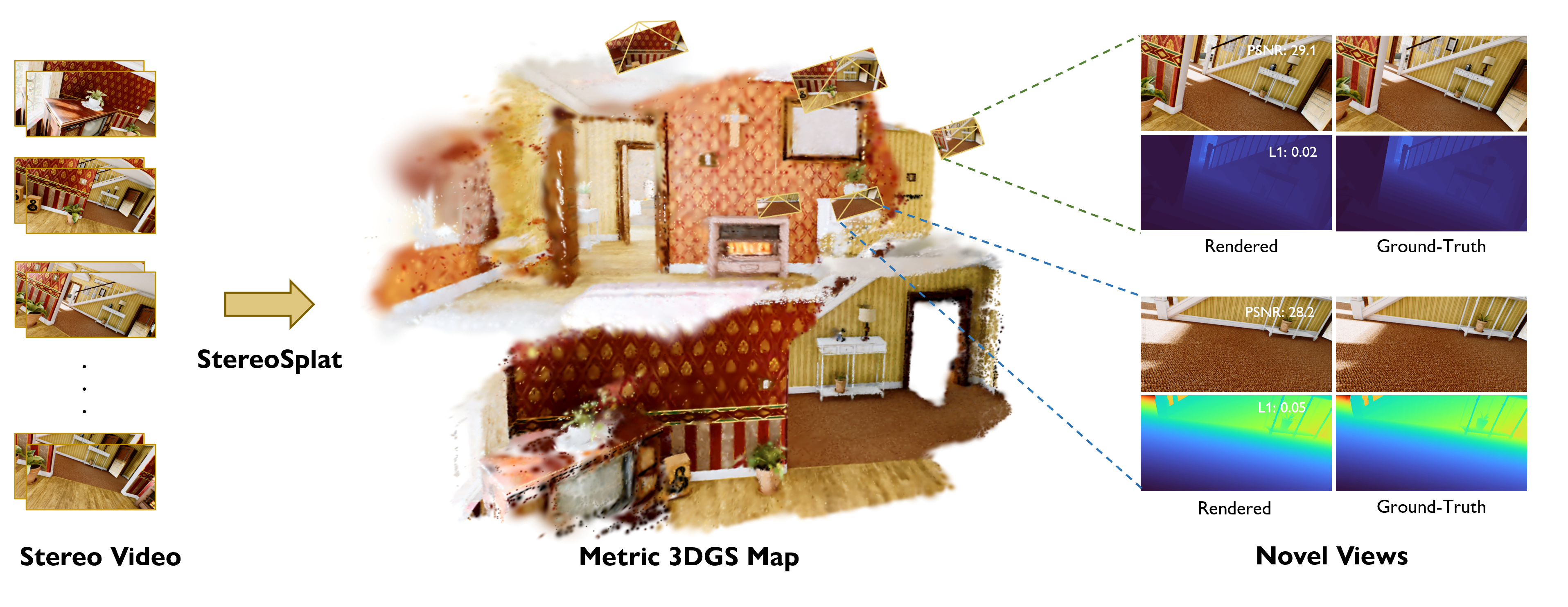}
  \caption{\ours is a feed-forward framework to reconstruct a 3D Gaussian Splatting (3DGS) representation from calibrated stereo videos. Our novel stereo-native architecture leverages a stereo prior, an effective decoder, and a robust 3DGS aggregation mechanism to directly predict low-redundancy 3D Gaussians in world-grounded metric space from input views.}
  \label{fig:teaser}
  \vspace{-1em}
\end{figure}
\begin{abstract}
We present \ours{}, a feed-forward 3D Gaussian Splatting architecture designed explicitly for stereo videos to achieve reliable, metric-scale novel-view synthesis. Unlike monocular approaches that neglect the nature of binocular pairs, \ours{} adopts a native stereo architecture that extracts dense multi-scale latent features and disparity from a foundation stereo backbone and feeds them into a dedicated Gaussian decoder. Our decoder utilizes cross-view feature fusion and gated depth refinement to directly translate these rich representations into per-pixel 3D Gaussians. To resolve the high redundancy of per-pixel predictions across overlapping views, we propose a learnable depth-adaptive aggregation mechanism that clusters and aggregates primitives in inverse-depth space while preserving fine-grained details. To support training and generalization, we introduce \oursdataset{}, a large-scale synthetic dataset of stereo video featuring various rig configurations tailored to match the physical hardware of real-world platforms. Trained exclusively on synthetic data and evaluated across challenging benchmarks, \ours{} yields substantial improvements in the depth accuracy of rendered novel views over established baselines while maintaining high-fidelity novel view synthesis. The model also transfers zero-shot to real-world stereo video, with applications in mixed reality, autonomous driving, and robotics. 
\end{abstract}

\section{Introduction}
\label{sec:intro}

The introduction of 3D Gaussian Splatting (3DGS)~\citep{kerbl2023gaussian_splatting} has established a new paradigm for real-time novel view synthesis, representing scenes as collections of anisotropic Gaussians that can be rendered at interactive frame rates. In parallel, scalable feed-forward architectures originating in natural language processing~\citep{vaswani2017transformer} and rapidly adopted in 2D~\citep{dosovitskiy2020vit} and 3D~\citep{wang2024dust3r,wang2025vggt} vision have enabled bypassing costly per-scene optimization entirely. Recently, works have begun to merge these two approaches, training large models that predict 3DGS representations directly from sparse input views in a single forward pass~\citep{charatan23pixelsplat,chen2024mvsplat,jiang2025anysplat}. The resulting feed-forward 3DGS methods offer a compelling combination of real-time rendering, strong generalization across diverse scenes, and dramatically reduced reconstruction time compared to traditional multi-stage pipelines.

Despite this progress, current feed-forward 3DGS methods suffer from the fundamental limitation of lacking solid geometric grounding. Methods that rely on learned cost volumes or cross-attention to implicitly infer depth, such as pixelSplat~\citep{charatan23pixelsplat} and MVSplat~\citep{chen2024mvsplat}, produce noisy geometry that is only recoverable up to an arbitrary scale. Even recent approaches incorporating foundational depth priors~\citep{xu2024depthsplat} or 3D models~\citep{ye2024noposplat,jiang2025anysplat} remain anchored to scale-ambiguous, affine-invariant representations. While the synthesized novel views might look plausible in color, the underlying geometry is often inaccurate, lacks fine detail, and degrades with distance. This limitation, compounded by the lack of metric scale, renders these methods unsuitable for downstream tasks requiring high geometric accuracy. A key reason is that these models treat inputs as generic, unstructured images. In doing so, they discard the calibration information and epipolar constraints that structured camera setups, such as stereo rigs, readily provide.

Stereo vision offers a principled solution to this geometric grounding problem. Given a calibrated stereo rig with known baseline and intrinsics, which is a standard setup in XR devices and robotics, dense absolute metric depth can be recovered directly from disparity via triangulation. The recent emergence of stereo matching foundation models, such as FoundationStereo~\citep{wen2025stereo}, has made this practical at scale by providing robust zero-shot disparity estimation across drastically different environments alongside rich multi-scale features that encode both local geometric detail and high-level scene semantics. However, to the best of our knowledge, no existing work bridges stereo foundation models with explicit 3DGS scene representations, and current architectures do not exploit stereo vision priors.

To address this gap, we present \ours, the first feed-forward 3D Gaussian Splatting framework designed explicitly for stereo videos. Rather than relying on generic image-to-3D backbones that ignore the inherent geometric constraints of stereo cameras, we leverage a stereo disparity foundation model as our core feature extractor. We further propose a decoding module that processes stereo features and uses the metric stereo depth as a geometric anchor, so that scene reconstruction and novel view synthesis from multi-view inputs stay in metric scale. Given the high redundancy in per-pixel predictions across overlapping stereo views, we introduce a learnable spatial compression module that jointly models spatial and appearance information, reducing the memory and rendering cost of dense per-pixel 3DGS predictions without sacrificing quality. Finally, to enable robust generalization, we introduce a large-scale synthetic pre-training dataset comprising diverse and highly challenging stereo video sequences. Our contributions can be summarized as follows:
\begin{itemize}
    \item \ours, the first feed-forward 3DGS model natively designed for stereo videos, achieving generalizable metric-scale novel view synthesis.
    \item A stereo-aware Gaussian decoder that fuses cross-view features with a gated depth refinement mechanism to regress geometrically accurate 3D Gaussians.
    \item A novel scene aggregation mechanism that integrates geometric and appearance cues with learned confidences to efficiently fuse per-pixel primitives.
    \item A challenging, large-scale synthetic dataset for stereo video pre-training, alongside extensive evaluations demonstrating that \ours yields a significant improvement in geometric accuracy over state-of-the-art feed-forward methods given the same stereo-video inputs.
\end{itemize}

\section{Related Work}
\boldparagraph{Foundational vision models.}
Since their introduction in machine translation~\citep{vaswani2017transformer}, transformers have become the dominant architecture in natural language processing~\citep{openai2023gpt4} and computer vision~\citep{he2022mae}. This trend has recently extended to 3D vision with the emergence of large-scale transformer models that jointly estimate camera poses and dense geometry~\citep{wang2025vggt, wang2025continuous}, or learn versatile representations for downstream 3D tasks~\citep{wang2024dust3r}. DUSt3R~\citep{wang2024dust3r} demonstrated that training transformer architectures on large-scale data using alternating self- and cross-attention enables the prediction of point maps in a shared coordinate system. VGGT~\citep{wang2025vggt} further extended this paradigm by scaling model capacity and predicting additional modalities beyond point maps, including tracks, depth, and camera poses. Despite their flexibility, these models are fundamentally constrained to unstructured, sparse-view, up-to-scale reconstructions and lack mechanisms for novel view synthesis. In contrast, our method operates directly on stereo sequences, enabling geometrically accurate, metric-scale reconstructions explicitly designed for high-fidelity novel view synthesis.

\boldparagraph{Feed-Forward Novel View Synthesis.}
While traditional NVS approaches require several optimization steps to reconstruct a scene via an explicit light field~\citep{levoy1996light, buehler2001unstructured} or an implicit neural field~\citep{mildenhall2021nerf, barron2022mip, muller2022instant}, generalizable methods train large-scale deep learning architectures to predict 3D scene representations in a single forward pass. Such approaches typically leverage neural rendering~\citep{yu2021pixelnerf, chen2021mvsnerf, flynn2024quark, jiang2025rayzer}, large reconstruction models~\citep{hong2024lrml, xu2024grm, gslrm2024, ziwen2025longlrm}, and 3D Gaussian Splatting (3DGS)~\citep{wang2024freesplat, charatan23pixelsplat, chen2024mvsplat, xu2024depthsplat}.
Among these works, 3DGS models are widely adopted for their real-time rendering. However, they frequently struggle to reconstruct the scene geometry and incur large memory footprints due to per-pixel Gaussian predictions. To reduce this overhead, AnySplat~\citep{jiang2025anysplat} combined the VGGT backbone with differentiable voxelization to compress the representation. Although this improves spatial efficiency, the underlying geometric reconstruction remains limited because depth is only predicted up to scale, leading to incorrect geometry when combined with unreliable pose estimation. In contrast, our method is explicitly designed to operate on stereo videos and exploits the metric depth of calibrated stereo rigs, resulting in higher-fidelity color rendering. Furthermore, we achieve these results while effectively compressing the per-pixel primitives to maintain memory efficiency. Concurrently with our work, TokenGS~\citep{tokengs2026} proposed compression by decoding splats from a fixed number of learnable tokens, but their framework is neither designed for stereo input nor compatible with dense depth priors.

\boldparagraph{Stereo matching.}
Deep stereo matching has progressed rapidly from early cost-volume architectures to large-scale foundation models. PSMNet~\citep{chang2018pyramid} introduced spatial pyramid pooling combined with 3D convolutions to regularize cost volumes, while GwcNet~\citep{guo2019group} improved correlation efficiency through group-wise feature matching. A pivotal shift came with the adoption of recurrent refinement. RAFT~\citep{teed2020raftrecurrentallpairsfield}, originally designed for optical flow, introduced an all-pairs correlation lookup with iterative GRU-based updates, and RAFT-Stereo~\citep{lipson2021raftstereomultilevelrecurrentfield} successfully adapted this paradigm to rectified stereo pairs. Subsequent methods further refined this iterative approach. For instance, IGEV-Stereo~\citep{xu2023iterative} combined geometry encoding volumes with iterative updates for improved accuracy, UniMatch~\citep{xu2023unifying} unified flow, stereo, and depth within a single framework, and Selective-Stereo~\citep{wang2024selective} introduced adaptive frequency selection to handle varying scene content. Most recently, FoundationStereo~\citep{wen2025stereo} scaled this iterative paradigm to a foundation model trained on massive and diverse data, achieving robust zero-shot generalization across domains while providing rich multi-scale features alongside disparity. In our work, we leverage FoundationStereo as a backbone, repurposing its metric output and intermediate feature representations as the foundation for feed-forward 3DGS prediction.

\section{Method}
\label{sec:method}
\begin{figure}[h]
  \centering
  \includegraphics[width=\linewidth]{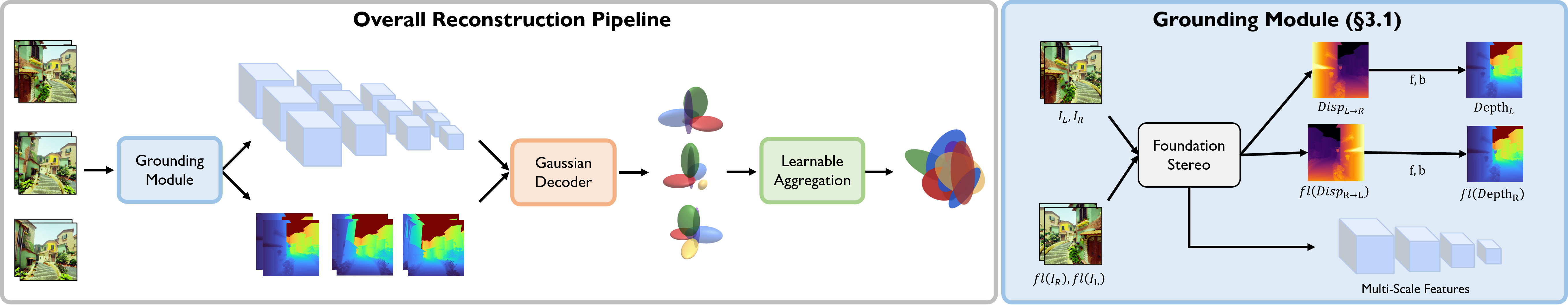}
  \includegraphics[width=\linewidth]{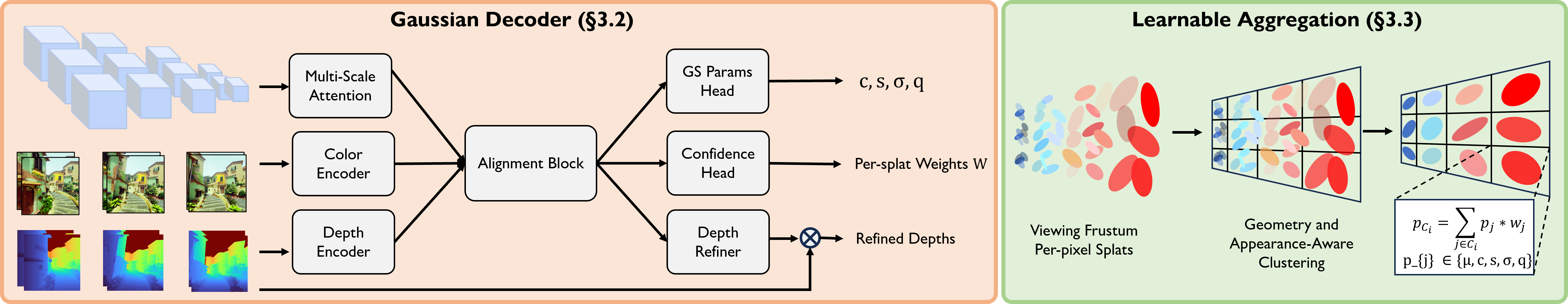}
  \vspace{-1em}
  \caption{Overview of our framework. \textbf{Top-left}: \ours consists of three key components: feature extraction and initial depth estimation, 3DGS decoding, and learnable primitive aggregation. \textbf{Top-right:} To extract per-view features and depth, the input stereo pair is flipped and reversed before passing through the backbone. Metric depth is then derived from the disparity maps using the camera focal length and stereo baseline. \textbf{Bottom-left:} Initial depth estimates and multi-scale features are aggregated in a shared feature space to regress 3DGS primitives and refine the depth.  \textbf{Bottom-right:} The aggregation module clusters per-pixel predictions based on disparity and appearance features. Finally, learned per-splat weights are used to fuse primitives within the same voxel.}
  \label{fig:architecture}
\end{figure}

Given a video sequence of $N$ rectified stereo image pairs $I = \{(I_i^{L}, I_i^{R}) \in \mathbb{R}^{H \times W \times 3}\}_{i=1}^{N}$, along with their corresponding camera extrinsics $T = \{(T_i^L, T_i^R) \in SE(3)\}_{i=1}^N$ and shared intrinsics $K$, our goal is to learn a mapping $f_\theta$ that directly predicts a set of 3D Gaussians $\mathcal{G}$ for novel view synthesis. Each Gaussian is parameterized by its mean in world coordinates $\mu \in \mathbb{R}^3$, appearance encoded via spherical harmonic coefficients $c \in \mathbb{R}^{3 \times (k+1)^2}$ of degree $k$, anisotropic scale $s \in \mathbb{R}_{\ge 0}^3$, opacity $\sigma \in [0, 1]$, and rotation represented as a unit quaternion $q \in \mathbb{H}$.

We design our feed-forward reconstruction framework around a robust geometric prior, leveraging a frozen stereo foundation model to anchor the scene geometry. Specifically, we extract dense metric depth and multi-scale context features via bidirectional stereo matching (Sec.~\ref{sec:grounding}). These features are processed by the decoder that regresses per-pixel Gaussian attributes. During decoding, the stereo-derived depth is refined via a gated residual connection before the pixels are unprojected into an initial, dense set of 3D primitives $\tilde{\mathcal{G}}$ (Sec.~\ref{sec:decoder}). Finally, to ensure rendering and memory efficiency, we compress this redundant set into a compact representation $\mathcal{G}$ using distance-adaptive clustering and learned attribute aggregation (Sec.~\ref{sec:aggregation}). The model is trained end-to-end via differentiable rendering using exclusively synthetic data, including the \oursdataset dataset (Sec. ~\ref{subsec:xrstereo}). An overview of our framework is illustrated in Fig.~\ref{fig:architecture}.

\subsection{Grounding module}
\label{sec:grounding}
For each stereo pair $(I_i^L, I_i^R)$, we use FoundationStereo \citep{wen2025stereo} as a frozen backbone $\mathcal{F}$ to extract initial depth and semantic features. Since $\mathcal{F}$ is designed to estimate disparity maps with respect to the left reference view ($d_i^L = \mathcal{F}(I_i^L, I_i^R)$), we exploit the symmetry of epipolar geometry to recover the corresponding right-view disparity. Specifically, we apply a horizontal reflection $\mathrm{flip}(\cdot)$ to the input pair to compute the right-view disparity:
\begin{equation}
d_i^R = \mathrm{flip}\!\Big(\mathcal{F}\big(\mathrm{flip}(I_i^R),\,\mathrm{flip}(I_i^L)\big)\Big)
\label{eq:flip}
\end{equation}
These disparities are subsequently projected into metric depth maps $D_i^v = f \cdot b / d_i^v$ for each view $v \in \{L, R\}$ using the focal length $f$ and baseline $b$ of the calibrated rig, which set the metric scale of reconstruction. In addition to depth, this bidirectional inference provides a consistent multi-scale context feature pyramid $\{\mathbf{F}_i^{v, s}\}_{s=1}^{S}$ for both input images in a single forward pass, which serves as the foundation for our method. 
\subsection{Gaussian decoder}\label{sec:decoder}
Using per-view images, depth maps, and feature pyramids, the decoder predicts a 3D Gaussian distribution for each pixel. This yields an initial set of primitives $\tilde{\mathcal{G}} = \{ \tilde{g}_{i}^{v} \}$ of size $|\tilde{\mathcal{G}}| = 2NHW$.
\paragraph{Cross-view feature fusion.}
To inject multi-view context, we apply a lightweight cross-view transformer at every scale without sharing weights. At each scale $s$ and spatial position $p$, we concatenate the left and right feature vectors across all views into a sequence of length $2N$, which is processed by a stack of pre-norm transformer blocks:
\begin{equation}
\hat{\mathbf{F}}^{(s)}(p) = \mathrm{Transformer}^{(s)}\!\left( \left[ \mathbf{F}_1^{L, s}(p), \mathbf{F}_1^{R, s}(p), \dots, \mathbf{F}_N^{L, s}(p), \mathbf{F}_N^{R, s}(p) \right] \right).
\label{eq:crossview}
\end{equation}
Each block comprises multi-head self-attention and a feed-forward network (FFN). Because spatial positions are processed independently, the attention matrix remains $2N \times 2N$, making the per-pixel computational complexity independent of the spatial dimensions $H_s \times W_s$. To ensure optimization stability, we zero-initialize the attention output projection and final FFN layer. This forces the module to initially act as an identity mapping, allowing it to smoothly adapt from the per-view baseline.
\paragraph{Gaussian Attribute Decoding.}
For brevity, we omit the pair index $i$ and view index $v$, as the following operations are applied identically to each input image. The decoder fuses full-resolution image features, multi-scale depth features extracted from $\log D$, and the cross-view enhanced pyramid $\{\hat{\mathbf{F}}^{(s)}\}$ into a single full-resolution tensor via a series of upsampling and residual blocks. Because the initial depth estimate $D$ can be inconsistent across views or time or because of flipping, we refine it by predicting a refined depth $\hat{D}$ and blending them using a learned scalar gate:
\begin{equation}
D' = (1-\alpha)\, D + \alpha\, \hat{D}, \qquad \alpha = \sigma(\alpha_0).
\label{eq:depth_refine}
\end{equation}
The learnable logit $\alpha_0$ is initialized to strongly bias the network toward the geometrically stable stereo estimate $D$ at the start of training. A linear head then maps each pixel feature to Gaussian attributes. Finally, using the refined depth $D'$, camera intrinsics $K$, and poses $T$, we unproject each pixel into world coordinates to define the 3D mean $\mu$, yielding the complete initial set of primitives $\tilde{\mathcal{G}}$.
\subsection{Gaussian aggregation}
\label{sec:aggregation}

The per-pixel set $\tilde{\mathcal{G}}$ is highly redundant. To reduce rendering costs, we compress $\tilde{\mathcal{G}}$ into a compact set $\mathcal{G}$ via hash-based aggregation. Because a Euclidean grid wastes capacity on distant geometry, we map Gaussian centers to inverse-depth coordinates $[x/z,\,y/z,\,1/z]$ and quantize them on a grid of size $\delta$. This naturally matches perspective projection. Lateral cell extent grows as $\delta z$ and depth extent as $\delta z^2$, preserving detail on nearby surfaces while aggressively consolidating distant primitives. We augment the hash key with quantized DC color, scale magnitude, and opacity to prevent mixing across discontinuities. Within each cluster $\mathcal{C}$, we aggregate attributes using learned importance weights. A linear head predicts a scalar logit $z_k$ for each primitive from its decoder features. We normalize these logits via a temperature-scaled softmax:
\begin{equation}
w_k = \frac{\exp(z_k / \tau)}{\sum_{j \in \mathcal{C}} \exp(z_j / \tau)}, \qquad \tau = \exp(\log\tau_0),
\label{eq:agg-weights}
\end{equation}
Here $\log\tau_0 \in \mathbb{R}$ is a learnable parameter controlling sharpness. By initializing $\log\tau_0 = 0$ to yield $\tau = 1$ and zero-initializing the weight head, all initial logits are zero. This reduces the softmax to a uniform distribution and exactly recovers standard \texttt{scatter\_mean} aggregation at the start of training. Over time, the network learns to up-weight informative primitives. Although cluster assignments are discrete, weighted aggregation remains differentiable, allowing gradients to flow back to the decoder.
\subsection{Training }
\label{sec:training}
We render the Gaussian scene $\mathcal{G}$ into target views using differentiable 3D Gaussian splatting~\citep{kerbl2023gaussian_splatting} via the \texttt{gsplat} rasterizer~\citep{ye2025gsplat}. Given a target camera with intrinsics $\mathbf{K}$ and pose $T \in SE_3$, the renderer projects and $\alpha$-composites the primitives to produce a predicted image $\hat{I}$ alongside a depth map $D_{\mathrm{rend}}$. We supervise the model against the ground-truth target image $I$ and its corresponding metric depth $D_{\mathrm{gt}}$ using a joint objective. The photometric loss relies on the mean absolute error, structural similarity, and perceptual metrics.
\begin{equation}
\mathcal{L}_{\mathrm{color}} =
\lambda_1 \lVert \hat{I}-I\rVert_1 +
\lambda_2 \bigl[1-\mathrm{SSIM}[\hat{I}, I]\bigr] +
\lambda_3 \mathrm{LPIPS}[\hat{I}, I]
\label{eq:color_loss}
\end{equation}
Here the scalars $\lambda_1, \lambda_2, \lambda_3 \in \mathbb{R}_{\ge 0}$ weight the respective terms. We concurrently apply a depth loss using the mean absolute error.
\begin{equation}
\mathcal{L}_{\mathrm{depth}} =
\lVert D_{\mathrm{rend}} - D_{\mathrm{gt}} \rVert_1
\label{eq:depth_loss}
\end{equation}
The complete training objective is $\mathcal{L}=\mathcal{L}_{\mathrm{color}}+\lambda_d \mathcal{L}_{\mathrm{depth}}$ with scalar weight $\lambda_d \in \mathbb{R}_{\ge 0}$. We optimize the decoder and aggregation modules end-to-end through the renderer.
\subsection{\oursdataset Dataset}
\label{subsec:xrstereo}
The robust generalization of feed-forward stereo reconstruction is currently limited by the scarcity of high-quality stereo video datasets that capture static environments. To address this critical gap, we introduce \oursdataset. This synthetic stereo video dataset comprises 137k high-quality frames split into 20k temporal sequences. These sequences feature moving cameras within static scenes across 9 indoor and 4 outdoor environments. To ensure visual diversity, each environment is populated with randomly placed objects sampled from a pool of 200 unique assets. All frames are rendered as rectified pinhole stereo pairs at 1K resolution with various rig configurations. These include horizontal fields of view ranging from 50 to 130 degrees and stereo baselines spanning 40 to 120 mm, reflecting the variability encountered in real VR/MR headsets and robotic platforms. Ground truth includes dense disparity maps, per-frame camera intrinsics, and extrinsics. A multi-stage quality assurance pipeline filters each frame to ensure sufficient non-dark pixels (>70\%), minimal near-plane clipping (<2\% of pixels below 11 cm depth), and low occlusion rates (>70\% non-occluded pixels), with sequence-level QA further requiring all five frames to individually pass these checks. Combined with TartanAir and Infinigen, \oursdataset provides the geometric and appearance diversity needed to train our model for zero-shot transfer to real-world stereo video.

\section{Experiments}
\boldparagraph{Datasets.} 
We train our method exclusively on synthetic data by combining complex samples from \oursdataset with the extreme camera trajectories of TartanAir~\citep{wang2020tartanairdatasetpushlimits} and the diverse stereo sequences of Infinigen~\citep{jing2024match} training set. We evaluate our approach across three distinct settings. First, we report novel view synthesis performance within the same domain using the unseen test split of Infinigen. Second, we evaluate generalization following the evaluation protocol established by~\citep{qiu2025mac} on TartanAir v2. This benchmark introduces significant rendering difficulties due to extreme viewpoint shifts and severe illumination changes. Finally, we evaluate real-world generalization using the KITTI~\citep{geiger2012kitti} dataset. This benchmark provides high-quality stereo videos and depth maps for testing our method outside the training distribution. Since our underlying 3D representation assumes a static scene, we manually exclude all video segments containing dynamic objects from both Infinigen and KITTI.

\boldparagraph{Evaluation Metrics.} We evaluate novel views by first analyzing geometric depth reconstruction, followed by color synthesis. To measure geometric consistency, we use the Absolute Mean Relative Error and $\delta_1$ accuracy following~\citep{jiang2025anysplat}. The Absolute Mean Relative Error calculates the average pixelwise normalized absolute depth difference, while the $\delta_1$ metric finds the percentage of pixels where the predicted to true depth ratio stays within an acceptable bound. We additionally report the L1 error for absolute depth estimation following~\citep{yugay2024magicslammultiagentgaussianglobally}. After validating geometric accuracy, we evaluate visual fidelity by computing PSNR, SSIM~\citep{wang2004image}, and LPIPS~\citep{zhang2018unreasonable} on full-resolution renders.

\boldparagraph{Baselines.} We evaluate \ours against state-of-the-art feed-forward novel-view synthesis methods spanning pose-aware and pose-free architectures. For pose-free baselines, we compare against AnySplat~\citep{jiang2025anysplat}, VicaSplat~\citep{li2025vicasplat}, and the Gaussian head of Depth Anything 3 (DA3-Giant)~\citep{depthanything3}. We exclude NoPoSplat~\citep{ye2024noposplat} because it strictly requires exactly two input images. Additionally, we benchmark against leading pose-aware models specifically DepthSplat~\citep{xu2024depthsplat}, MVSplat~\citep{chen2024mvsplat}, and PixelSplat~\citep{charatan23pixelsplat}. At the time of submission, the checkpoints for the concurrent work, TokenGS~\citep{tokengs2026}, have not been released.

\boldparagraph{Evaluation Protocol.} 
Following \citep{ye2024noposplat}, we mask out regions unseen by the input cameras. We adopt the evaluation protocol of \citep{xu20254dgt} for video sequences, dividing each video into local temporal windows of three stereo frames. We select this compact window to evaluate local temporal consistency, while extensive evaluations on larger context windows are provided in the supplementary material. Within each window, we select every other frame to evaluate novel view synthesis. All baselines receive the same input images as our method and are evaluated with their publicly released checkpoints, without retraining on stereo video. The pose-aware baselines MVSplat, PixelSplat, and DepthSplat use ground-truth poses. Because pose-free models inherently couple reconstruction and pose prediction in a normalized space, making ground-truth pose injection impossible, we follow the specific evaluation protocol introduced by \citep{ye2024noposplat, jiang2025anysplat} for these methods. Since DA3 predicts depth only up to scale, we additionally align its rendered depth to the ground truth with an oracle median scaling before computing geometry metrics, which leaves its appearance metrics unchanged. Furthermore, to evaluate each model within its intended operational limits, MVSplat, PixelSplat, and VicaSplat are tested at their natively supported $256 \times 256$ resolution, whereas \ours operates at $320 \times 640$. To simulate real-world deployment, we evaluate our method across all experiments using trajectories from an off-the-shelf visual odometry system \citep{qiu2025mac}.

Our framework is not tied to a specific stereo backbone. We provide implementation details, a runtime and memory analysis, results with an alternative stereo backbone, an ablation of the gated depth refinement, evaluations with different numbers of input frames, additional qualitative results, and ablations on \oursdataset in the supplementary material.
\subsection{Novel View Synthesis}
\label{sec:experiment}
\begin{figure*}[ht!]
  \centering
  \newcommand{\wratio}{0.24} 

  \makebox[0.01\textwidth][c]{}
  \makebox[\wratio\textwidth]{\normalsize AnySplat}
  \makebox[\wratio\textwidth]{\normalsize DepthSplat$^\dagger$}
  \makebox[\wratio\textwidth]{\normalsize \ours (Ours)}
  \makebox[\wratio\textwidth]{\normalsize Ground Truth}
  \\
  \raisebox{.4cm}{\makebox[0.01\textwidth]{\rotatebox{90}{\normalsize Color}}}
  \includegraphics[width=\wratio\textwidth]{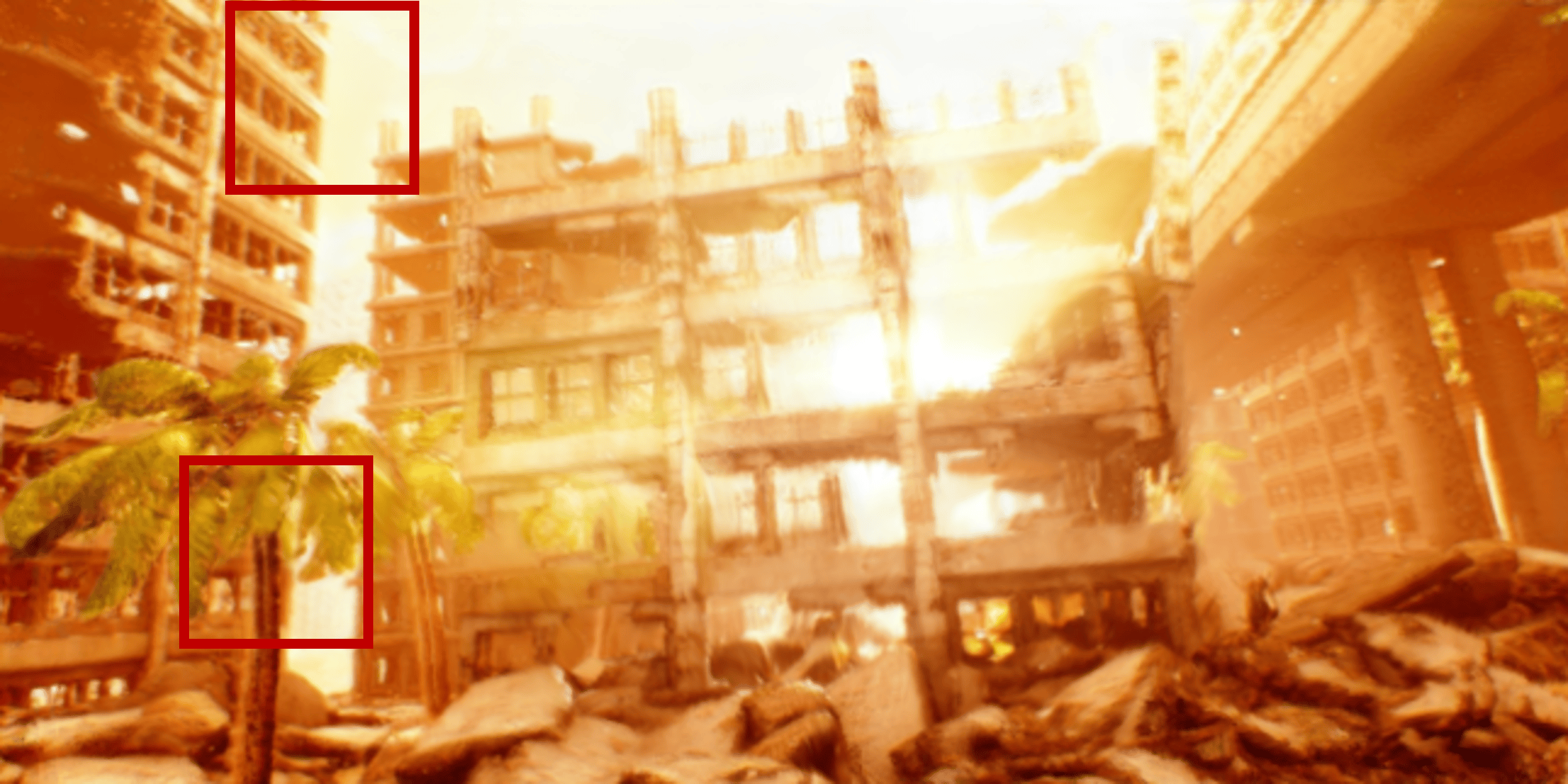}
  \includegraphics[width=\wratio\textwidth]{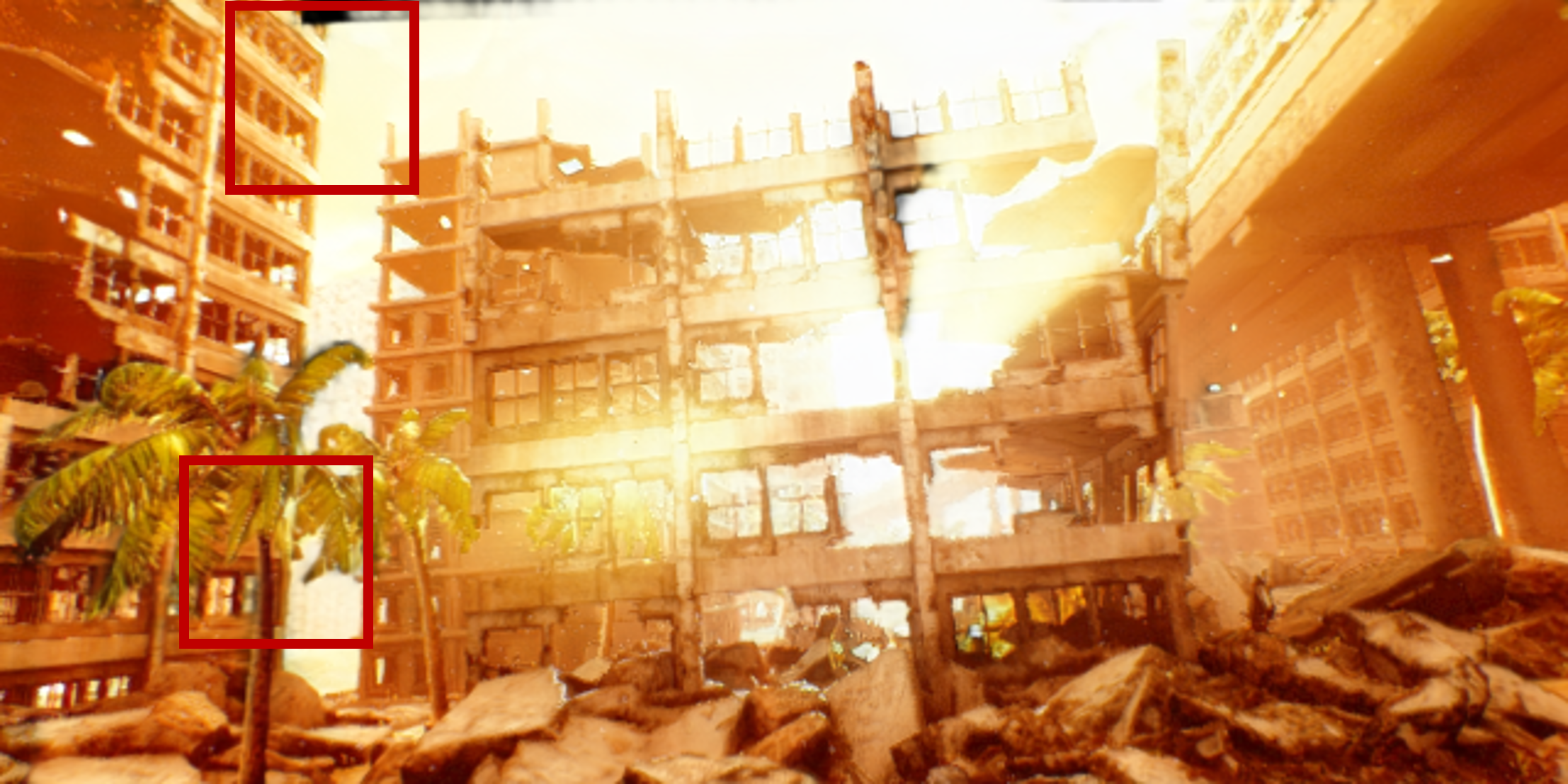}
  \includegraphics[width=\wratio\textwidth]{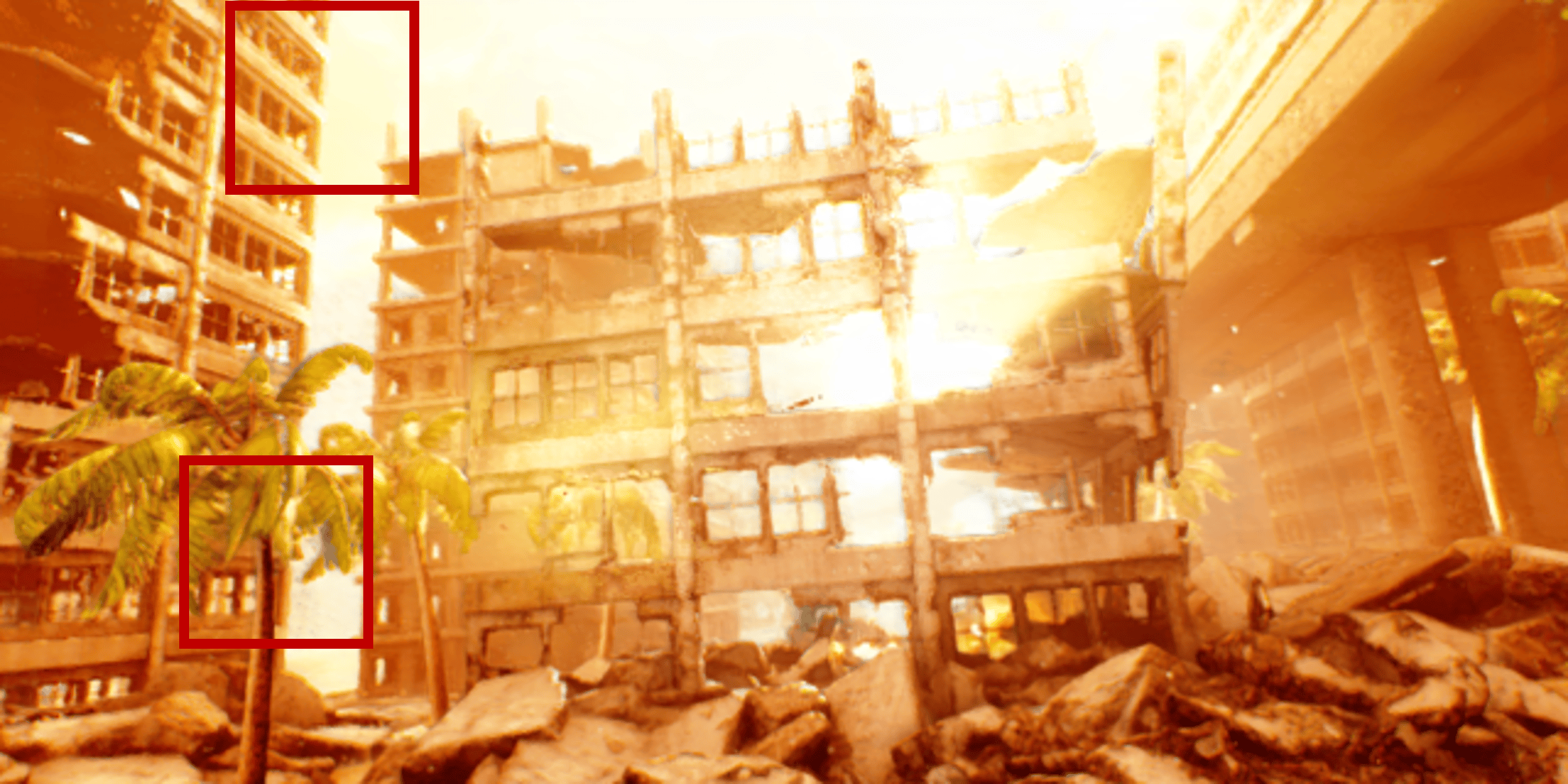}
  \includegraphics[width=\wratio\textwidth]{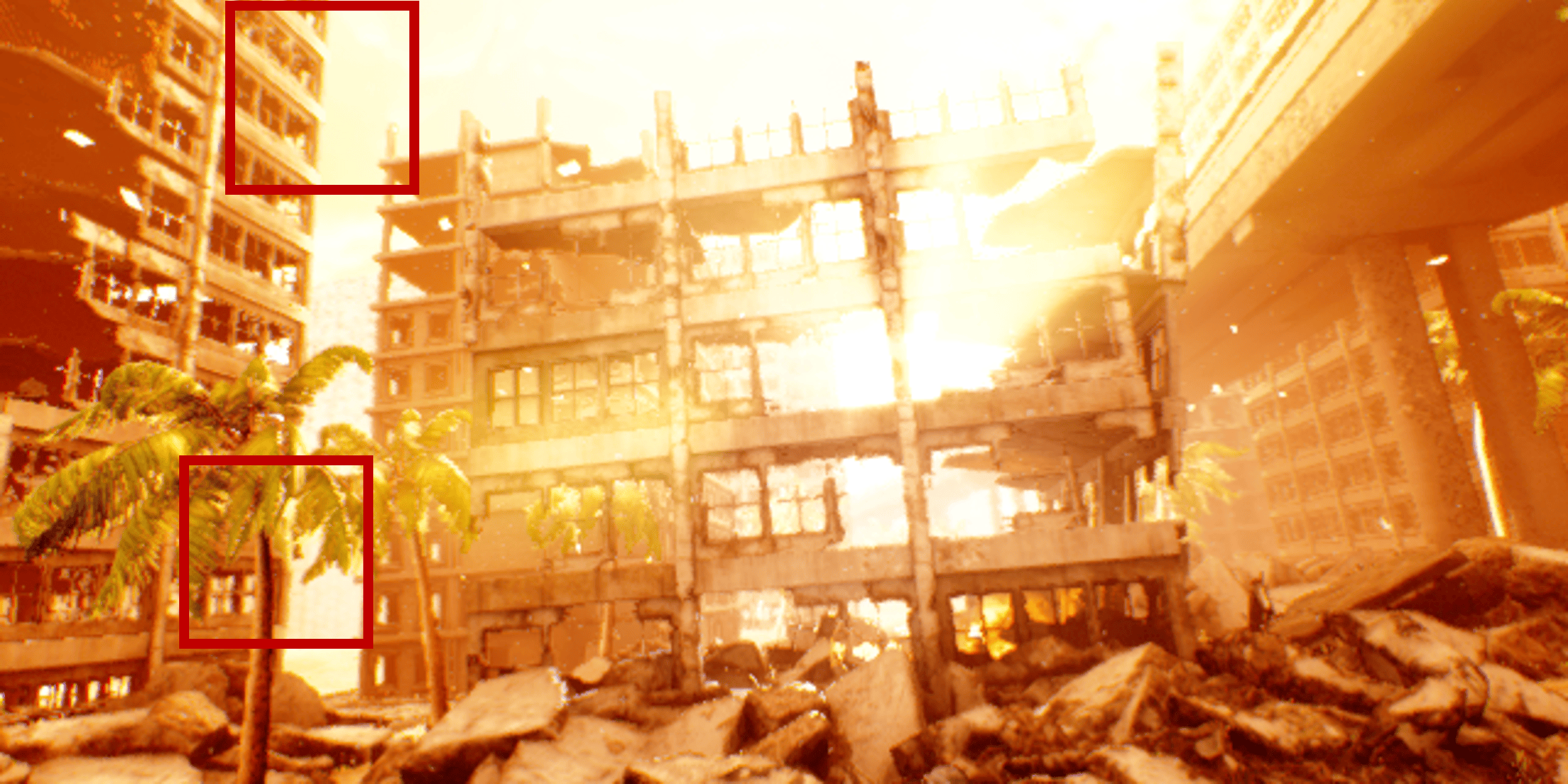}
  \\
  \raisebox{.4cm}{\makebox[0.01\textwidth]{\rotatebox{90}{\normalsize Depth}}}
  \includegraphics[width=\wratio\textwidth]{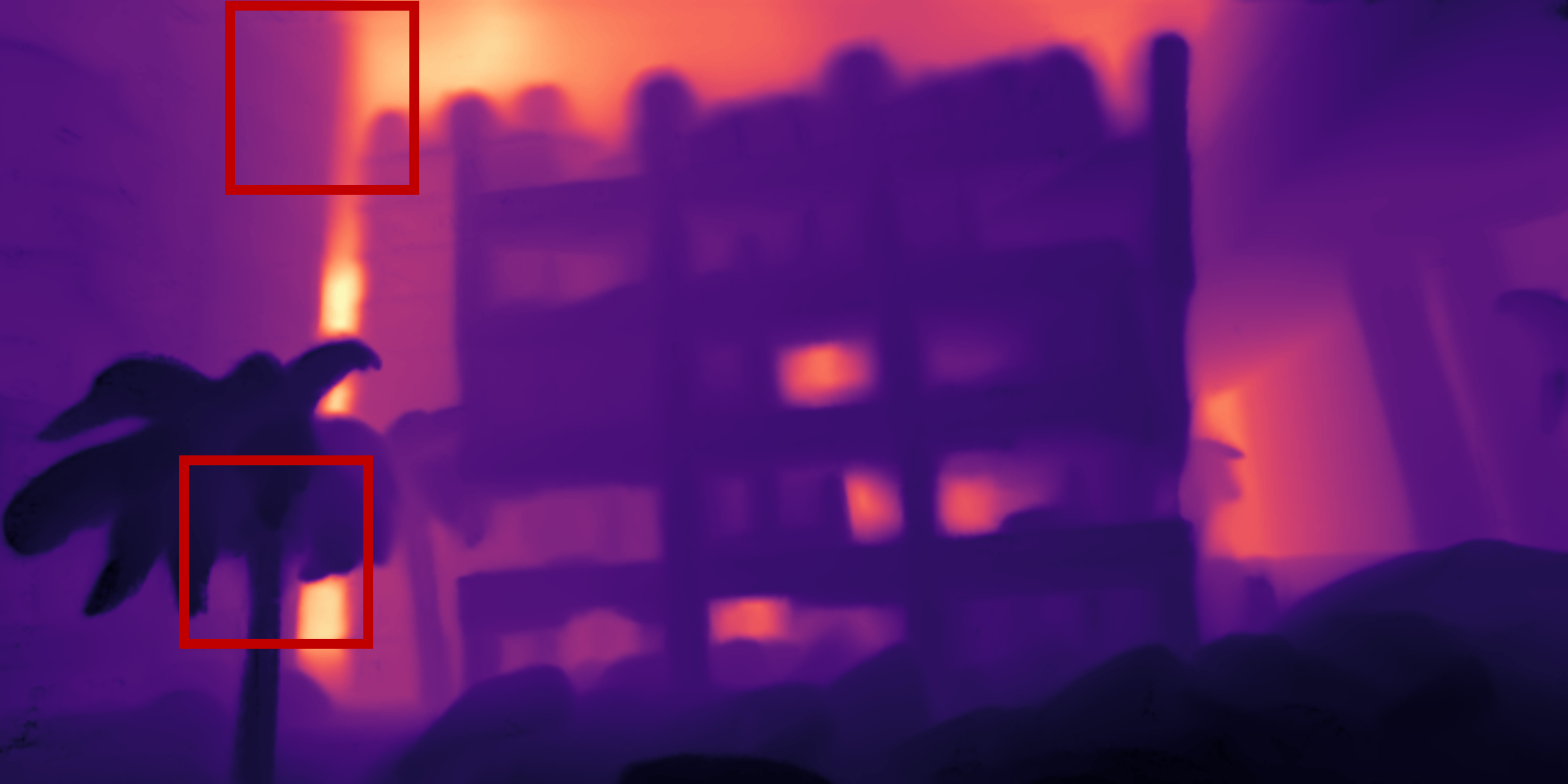}
  \includegraphics[width=\wratio\textwidth]{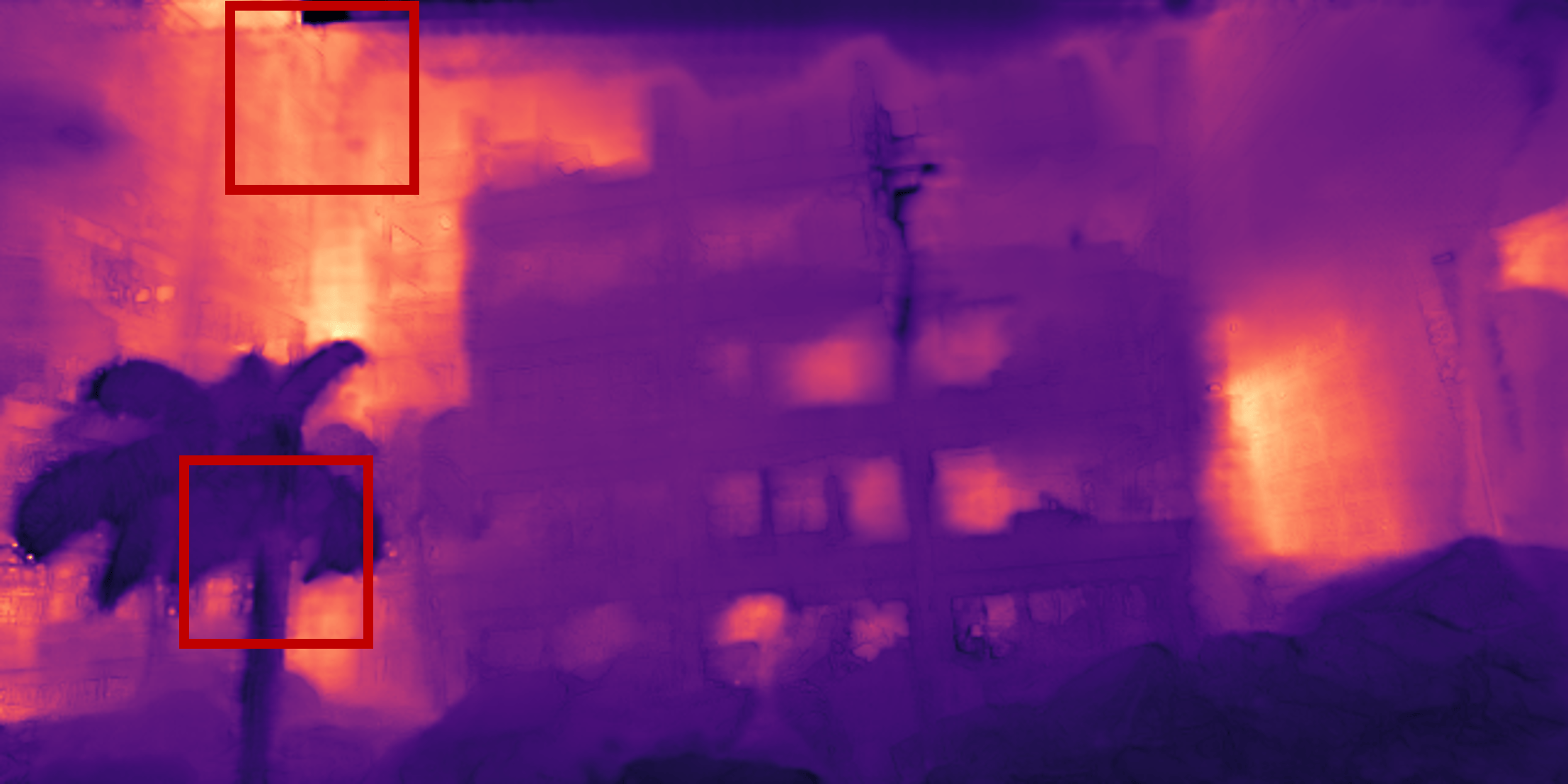}
  \includegraphics[width=\wratio\textwidth]{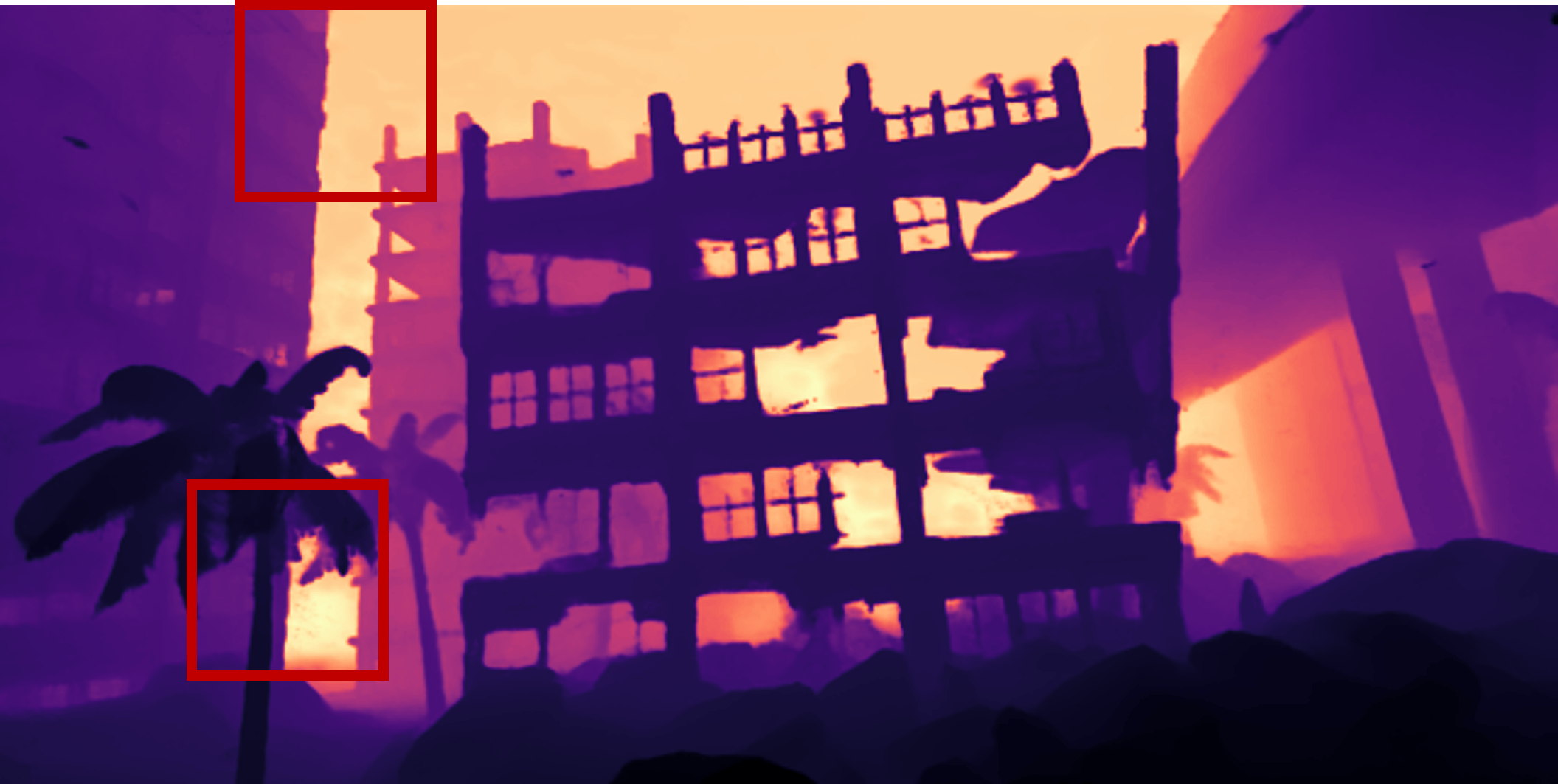}
  \includegraphics[width=\wratio\textwidth]{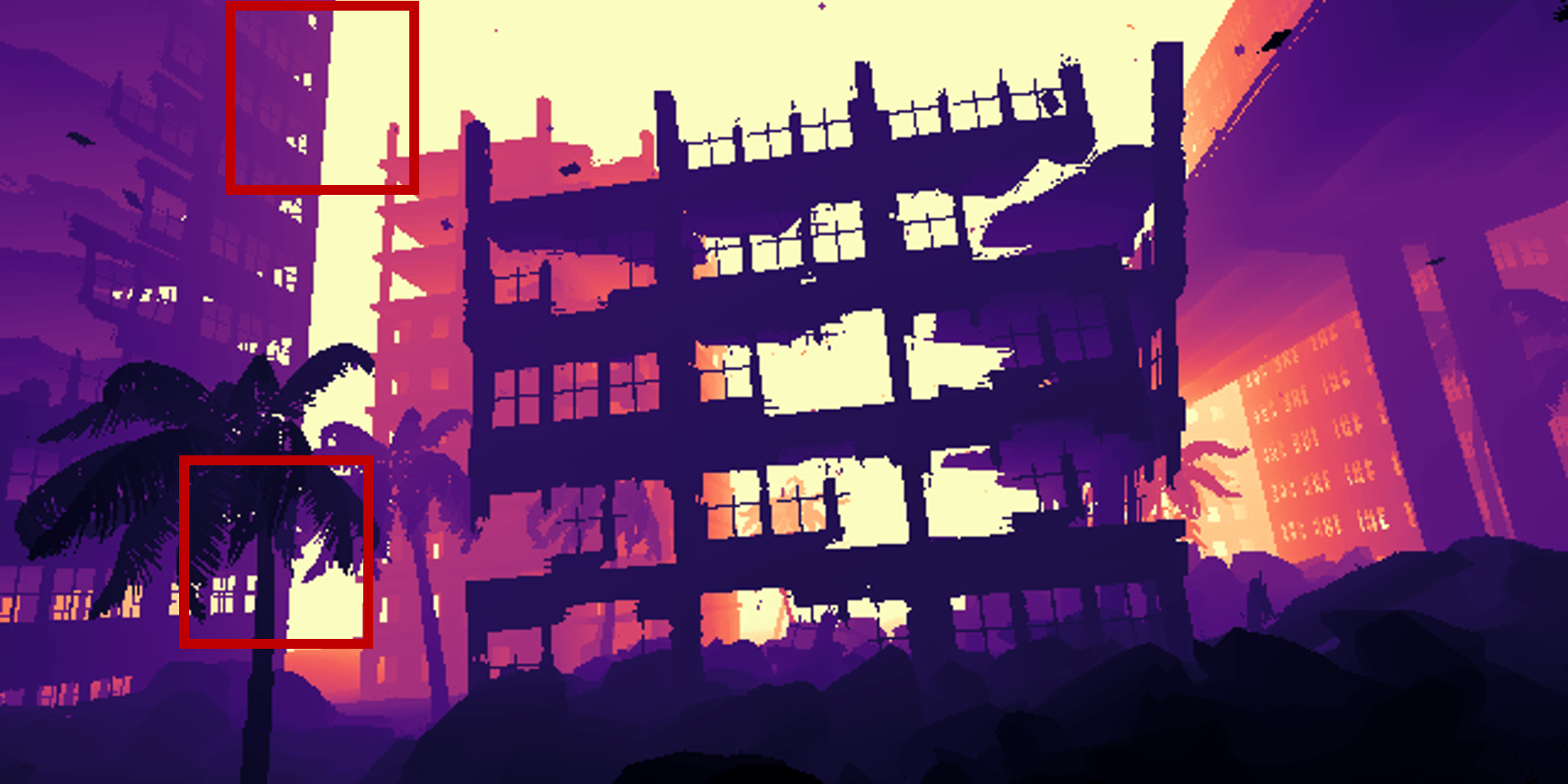} 
  \\
  \raisebox{.4cm}{\makebox[0.01\textwidth]{\rotatebox{90}{\normalsize Color}}}
  \includegraphics[width=\wratio\textwidth]{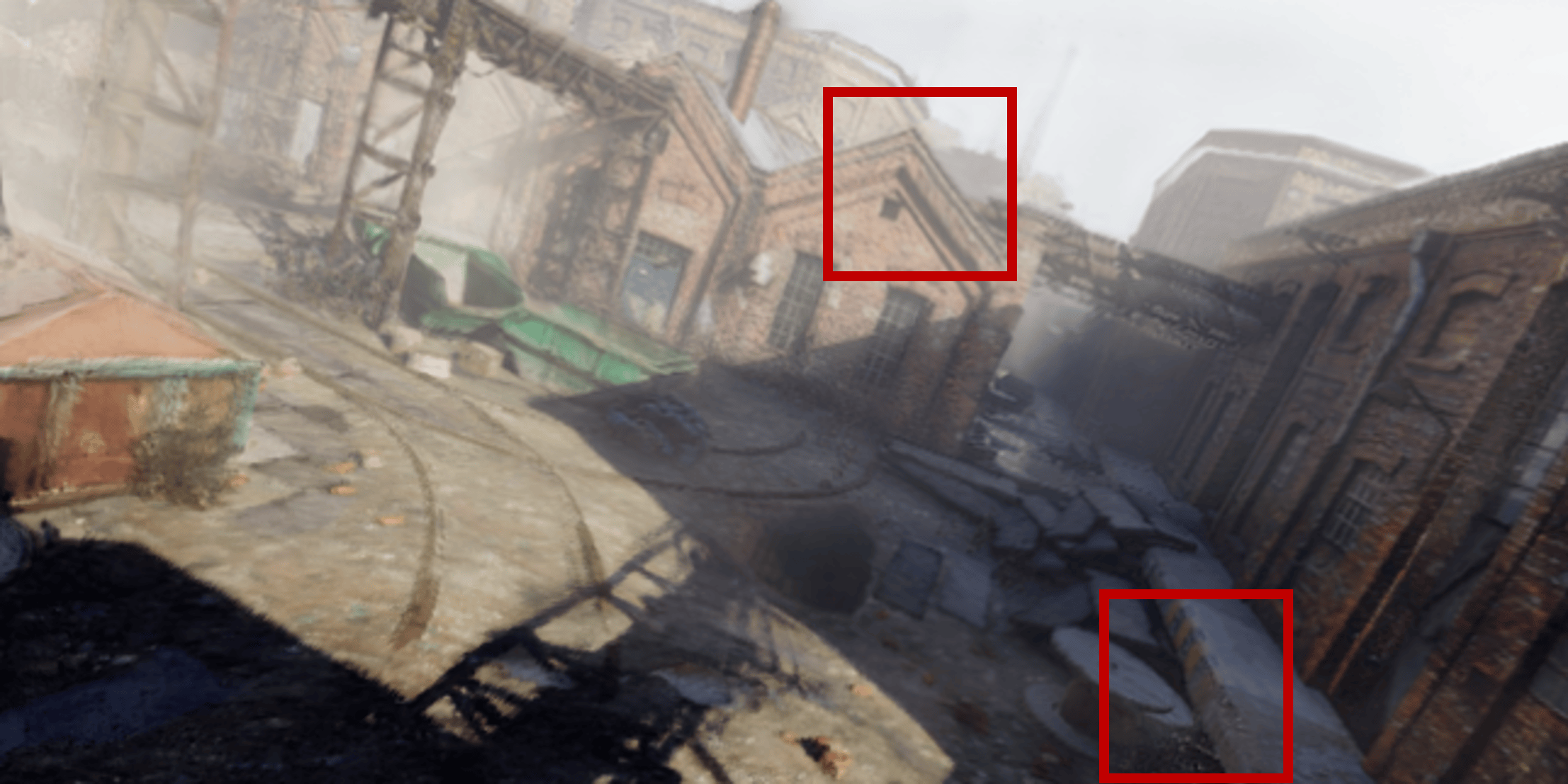}
  \includegraphics[width=\wratio\textwidth]{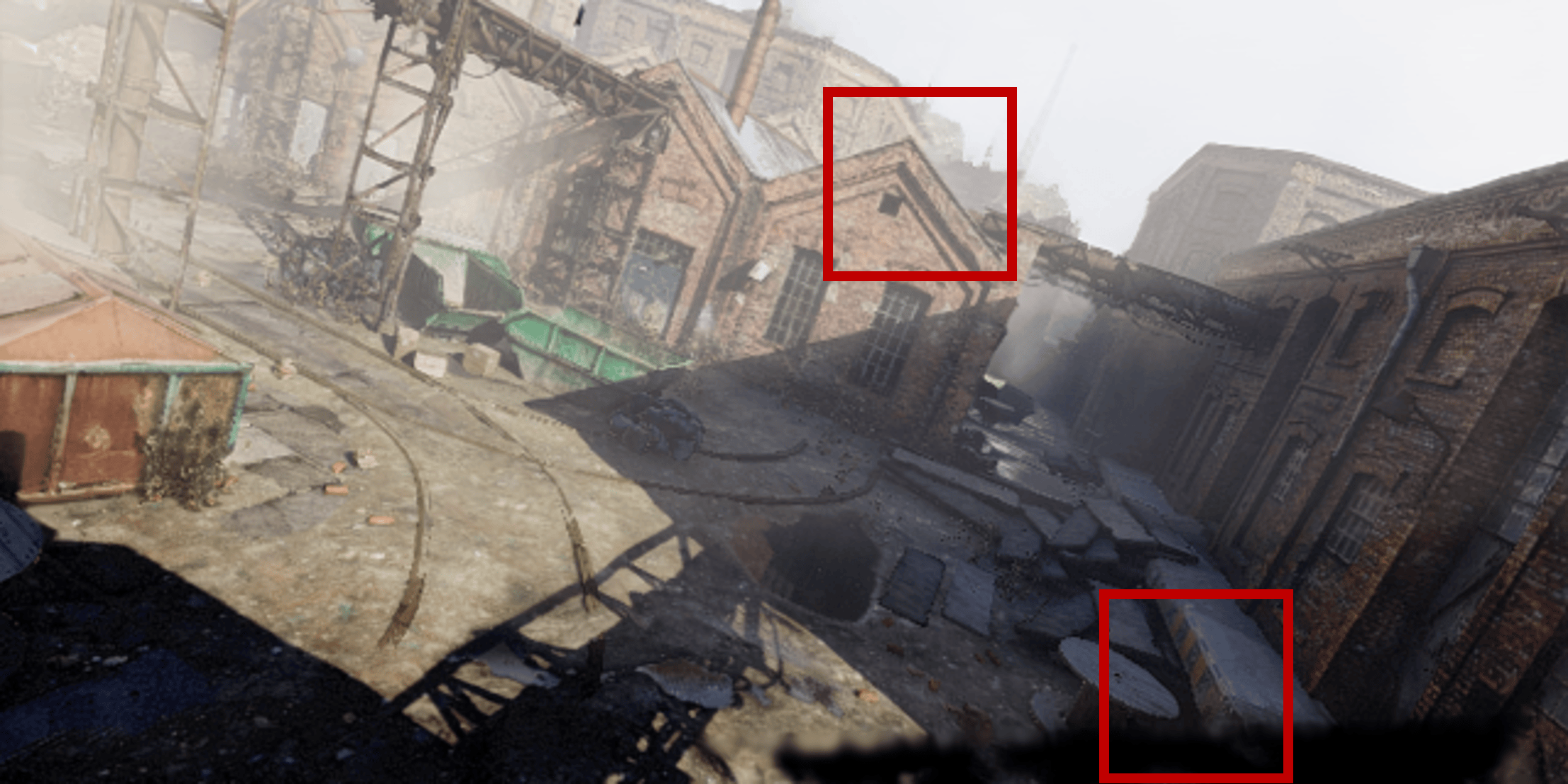}
  \includegraphics[width=\wratio\textwidth]{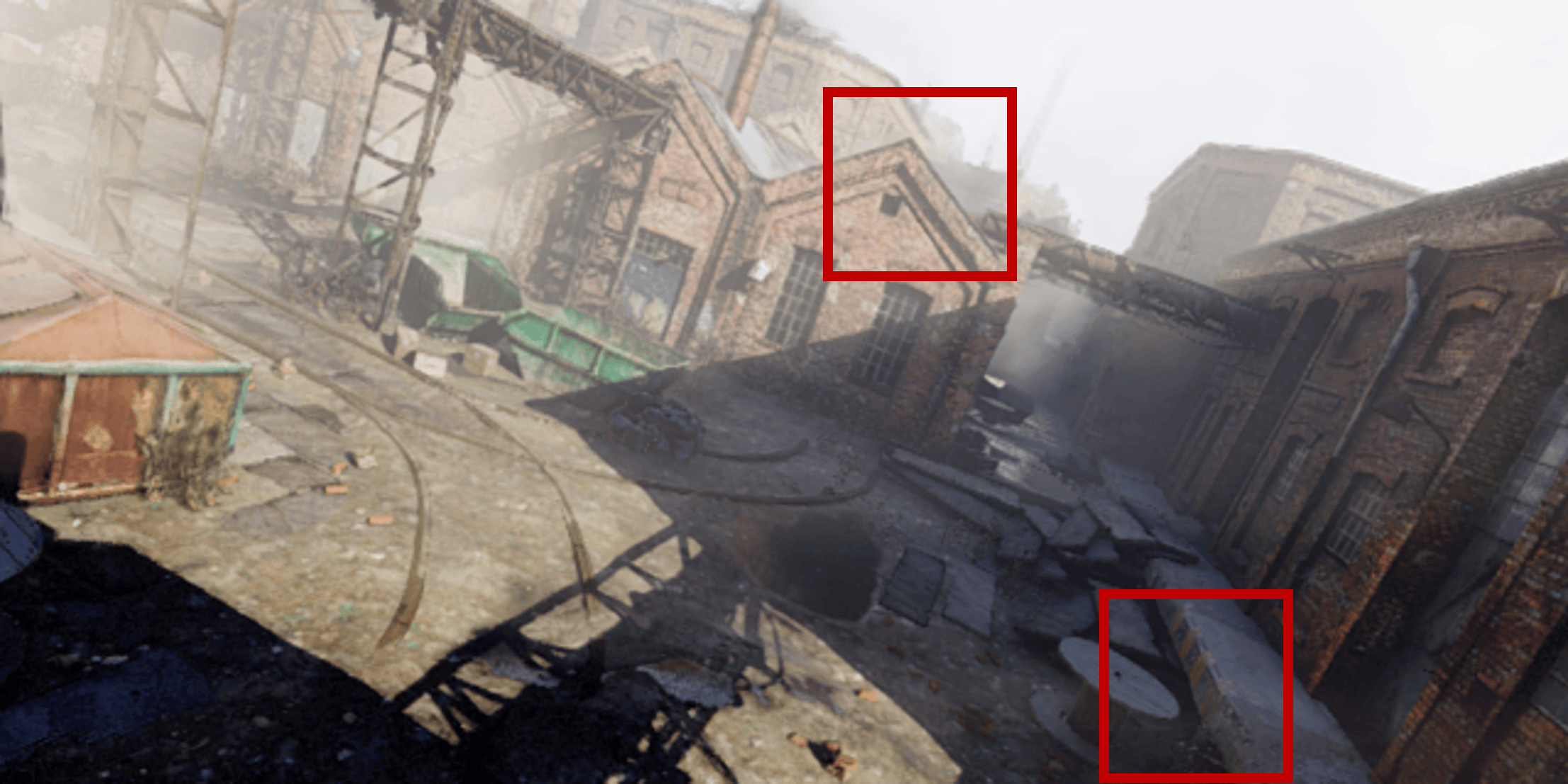}
  \includegraphics[width=\wratio\textwidth]{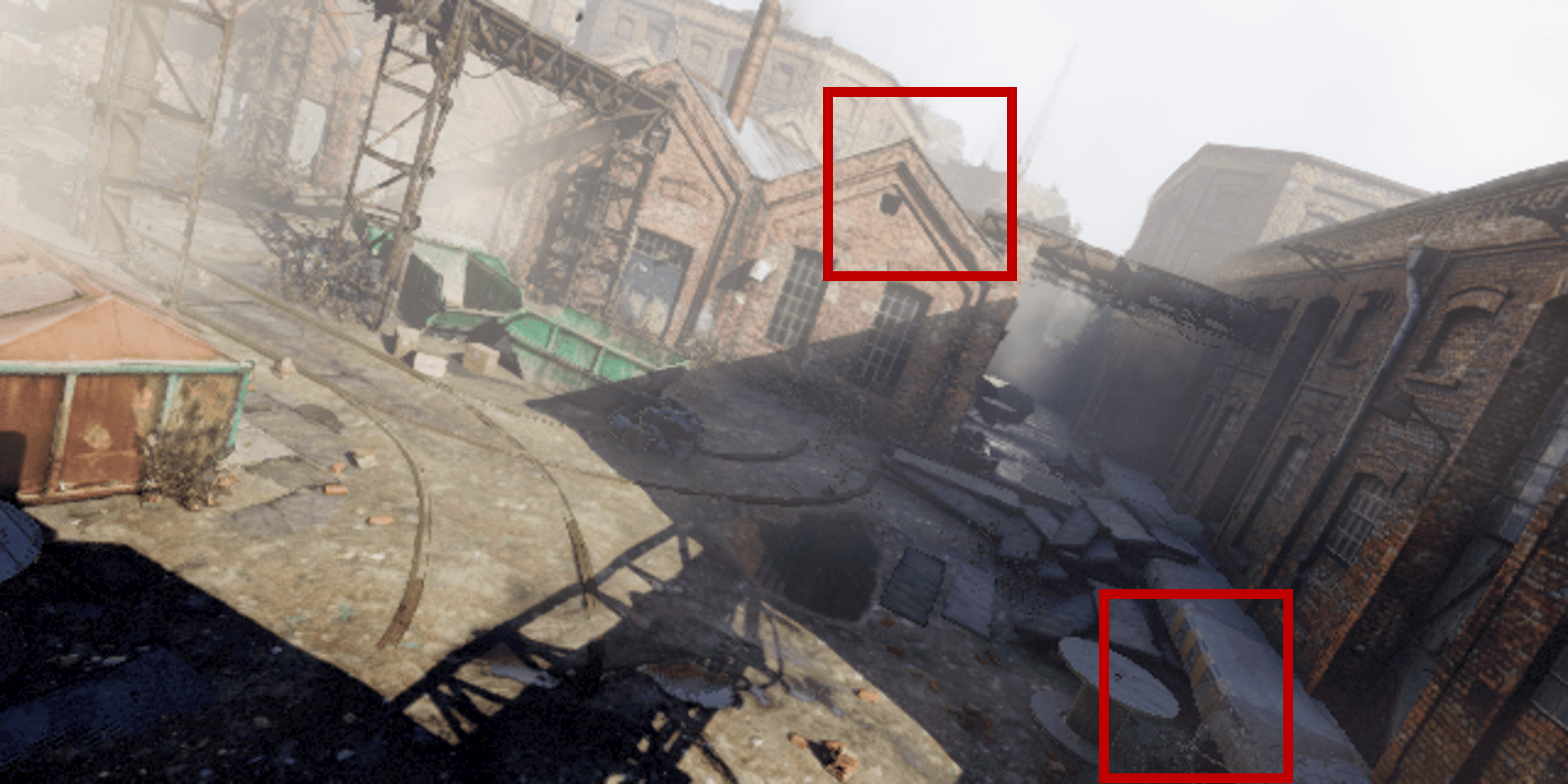}
  \\
  \raisebox{.4cm}{\makebox[0.01\textwidth]{\rotatebox{90}{\normalsize Depth}}}
  \includegraphics[width=\wratio\textwidth]{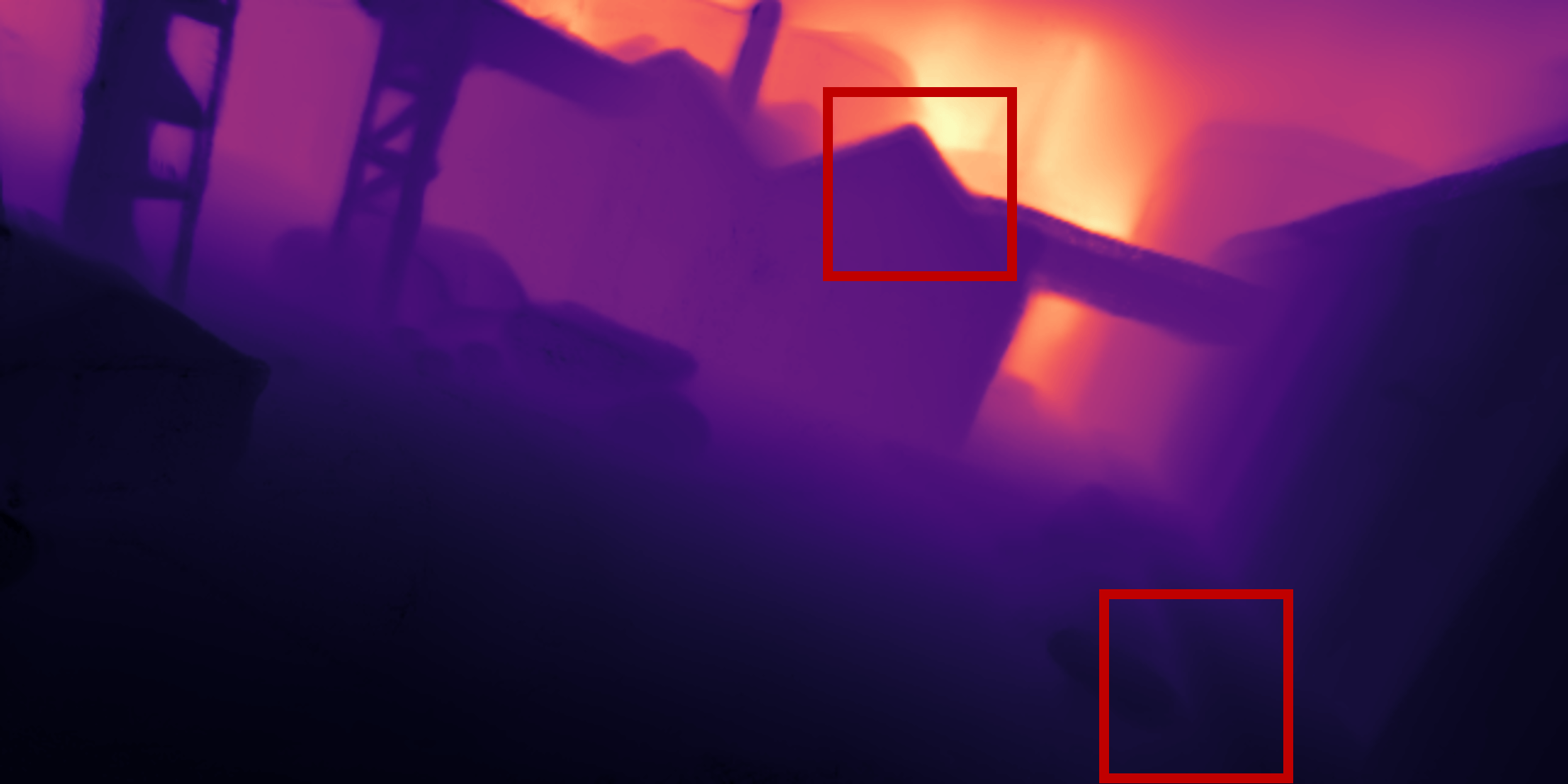}
  \includegraphics[width=\wratio\textwidth]{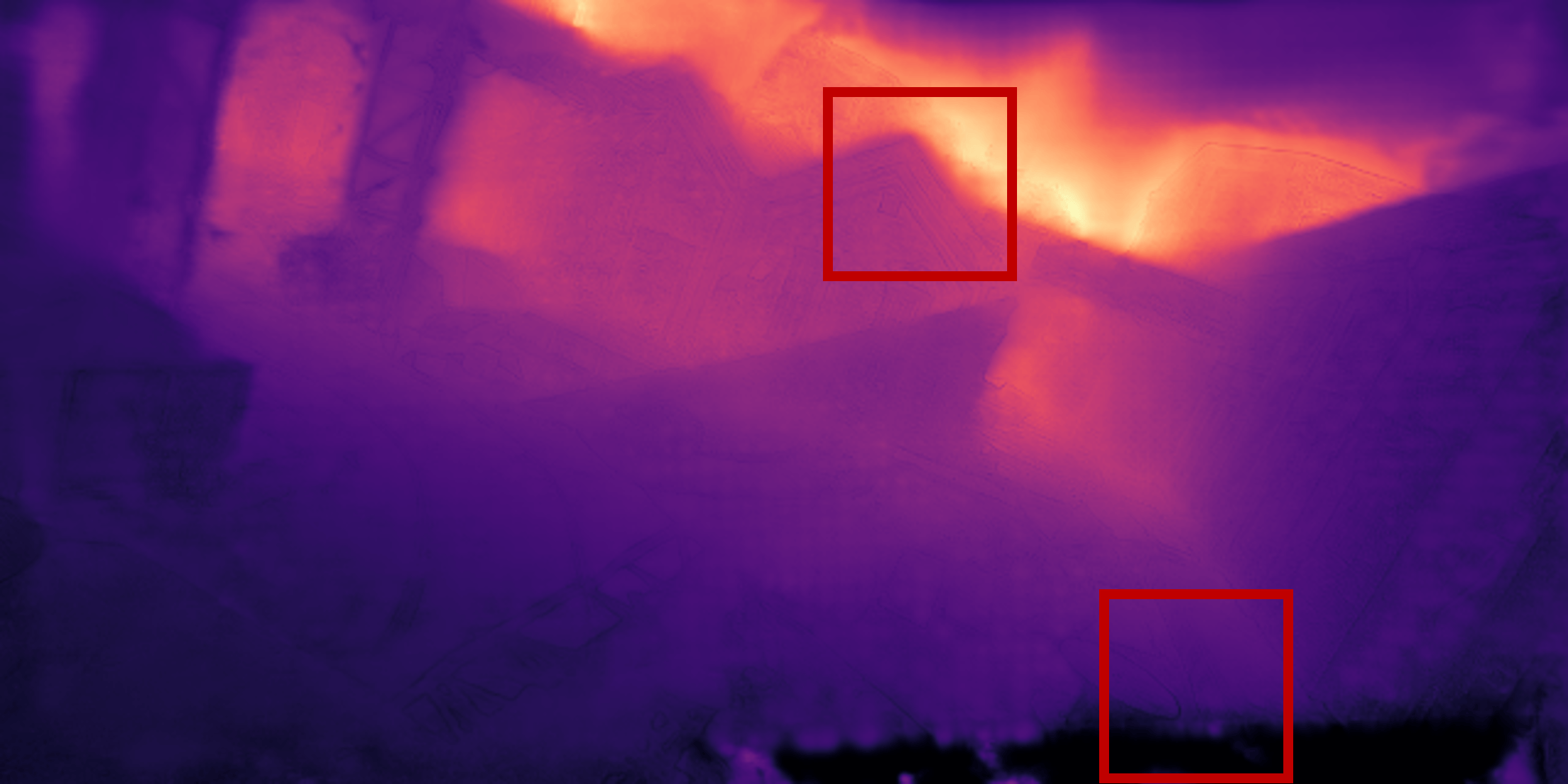}
  \includegraphics[width=\wratio\textwidth]{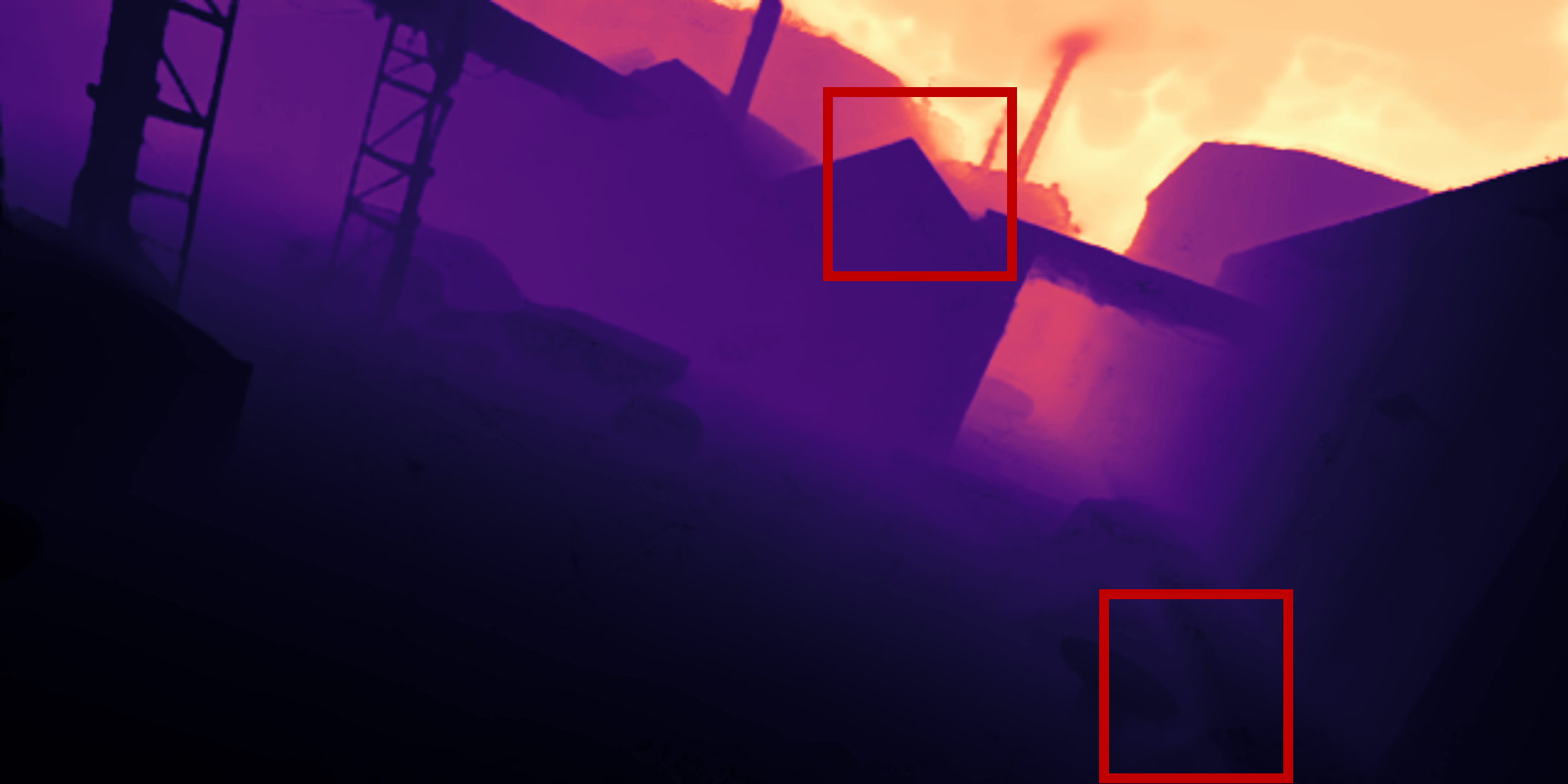}
  \includegraphics[width=\wratio\textwidth]{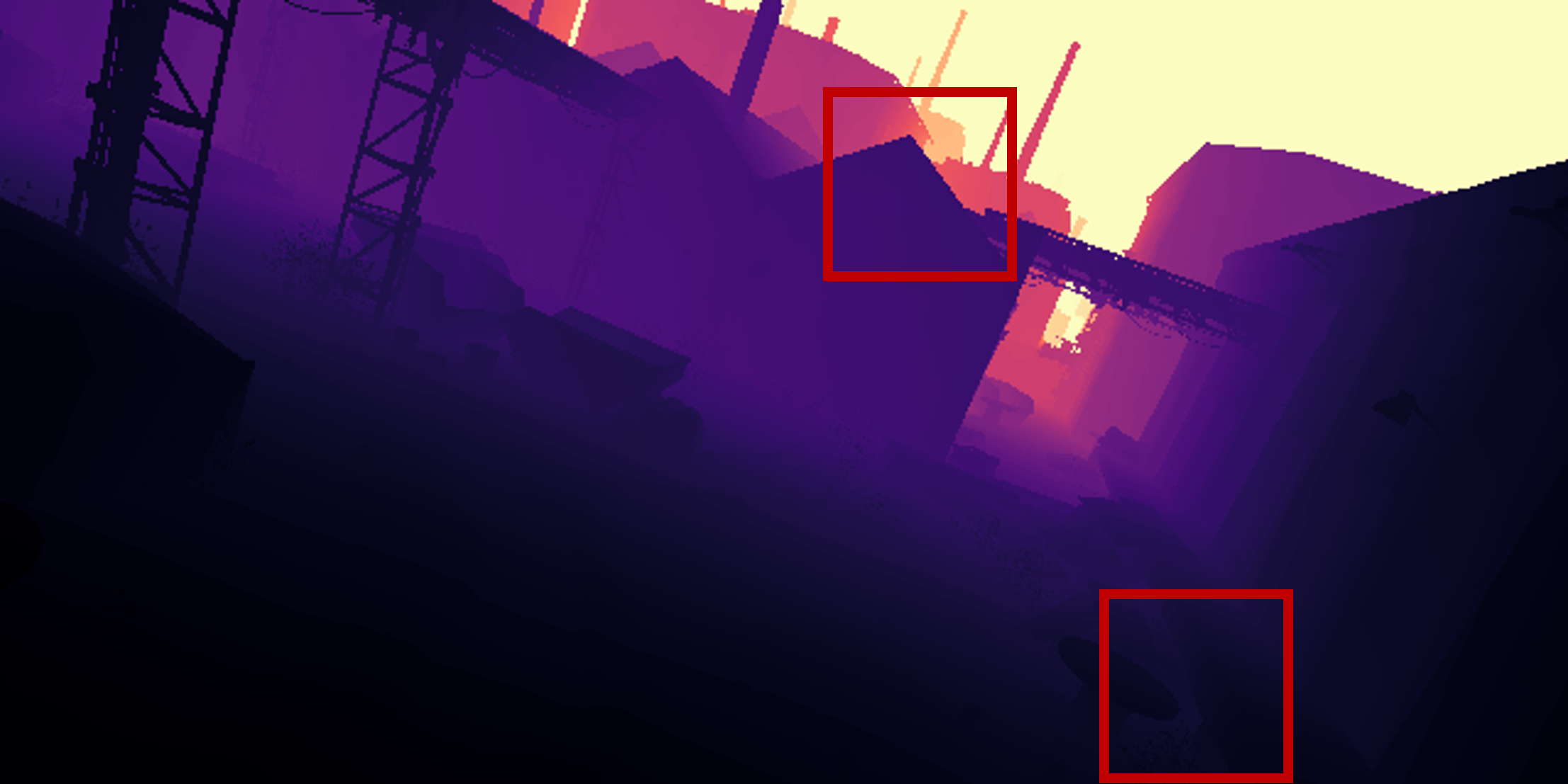}  
  \\
  \raisebox{.4cm}{\makebox[0.01\textwidth]{\rotatebox{90}{\normalsize Color}}}
  \includegraphics[width=\wratio\textwidth]{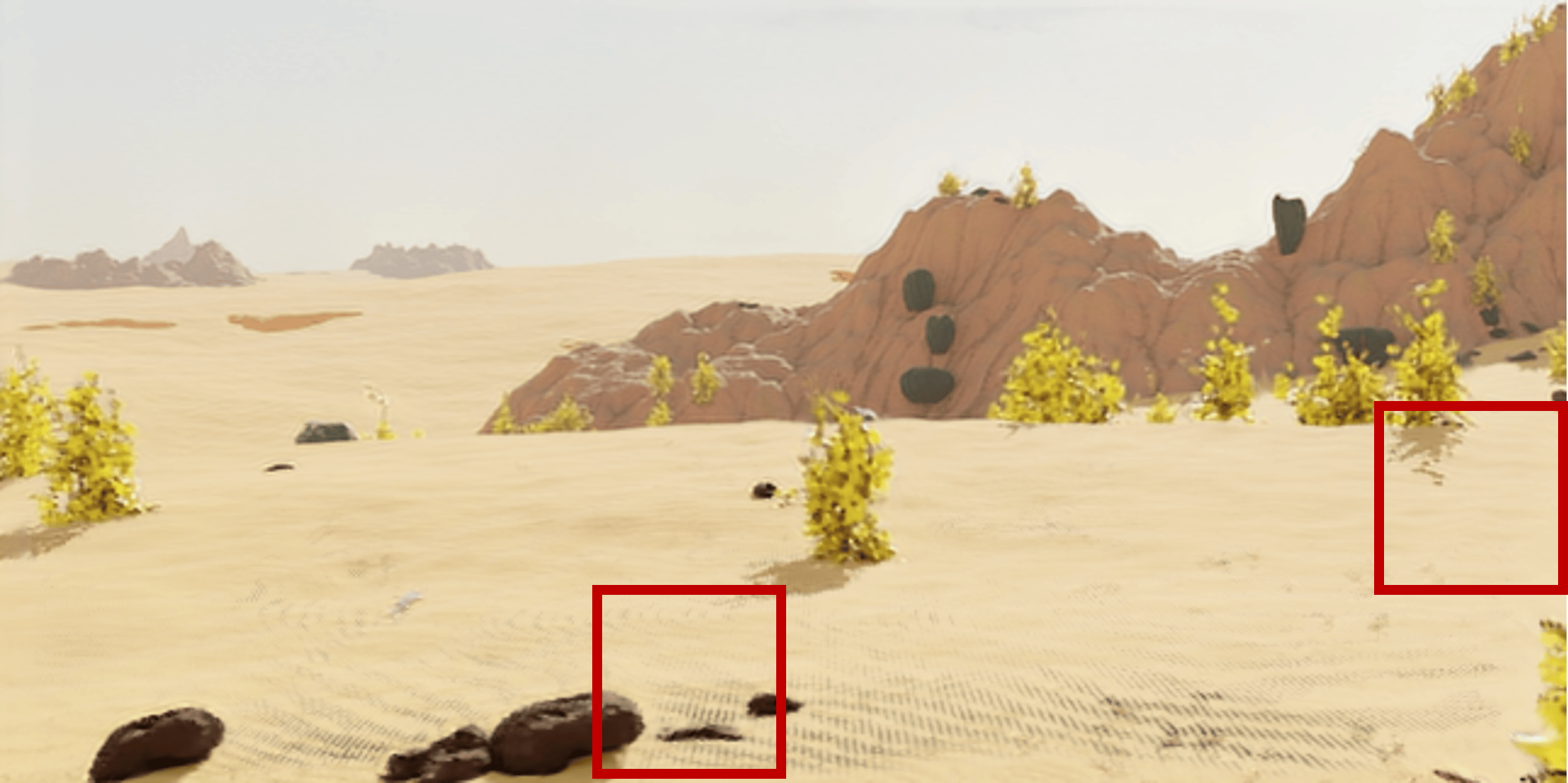}
  \includegraphics[width=\wratio\textwidth]{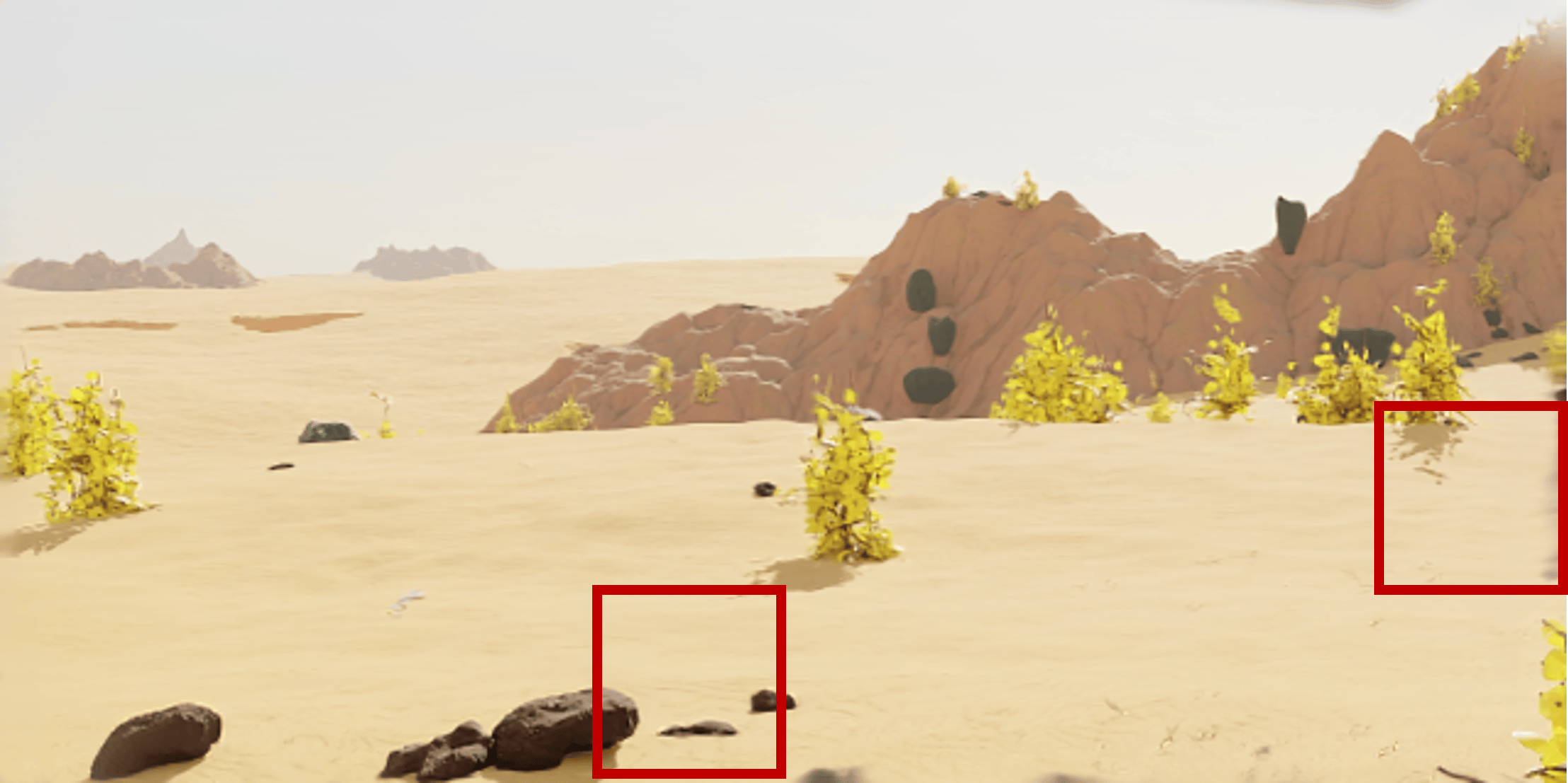}
  \includegraphics[width=\wratio\textwidth]{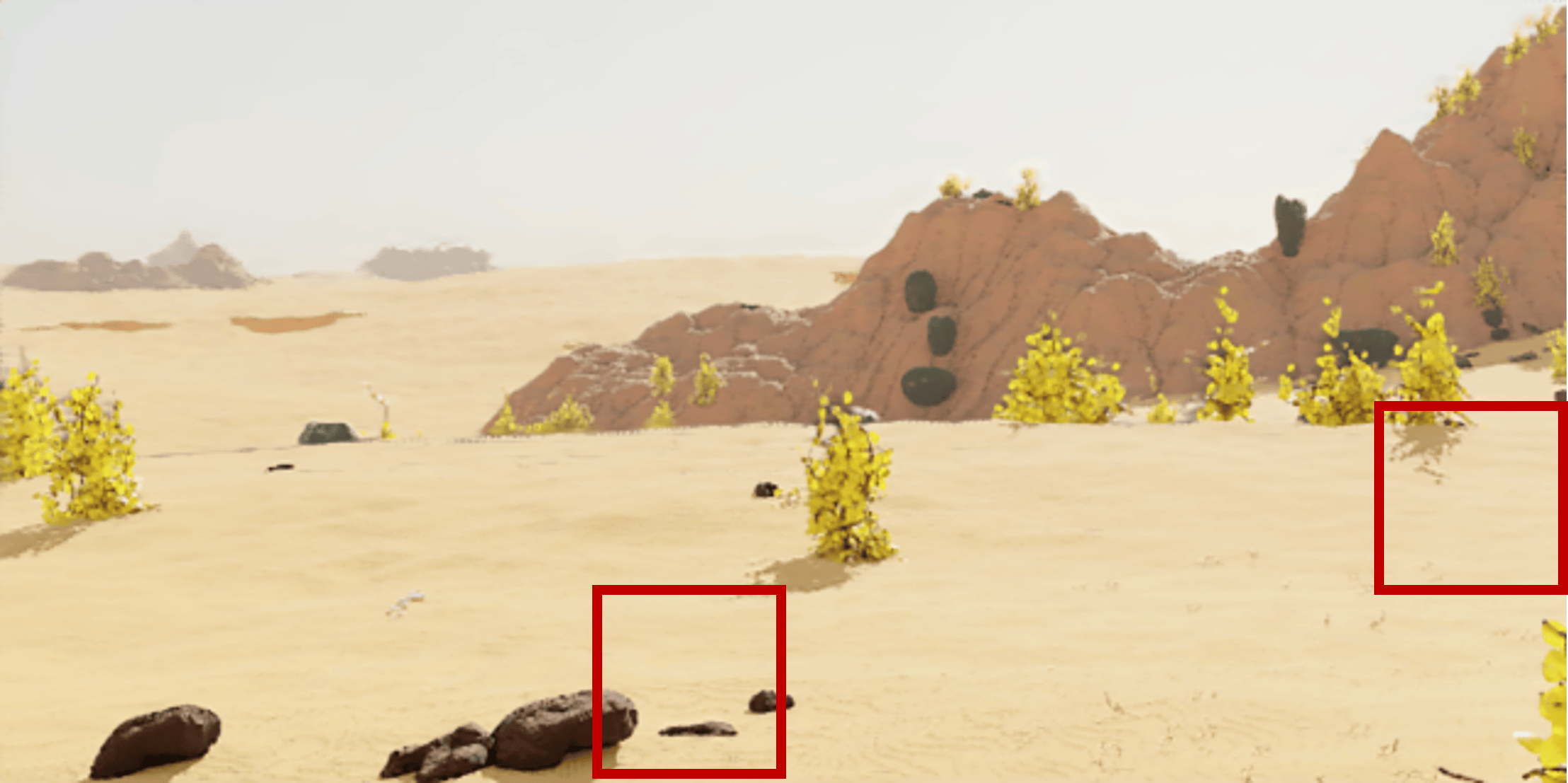}
  \includegraphics[width=\wratio\textwidth]{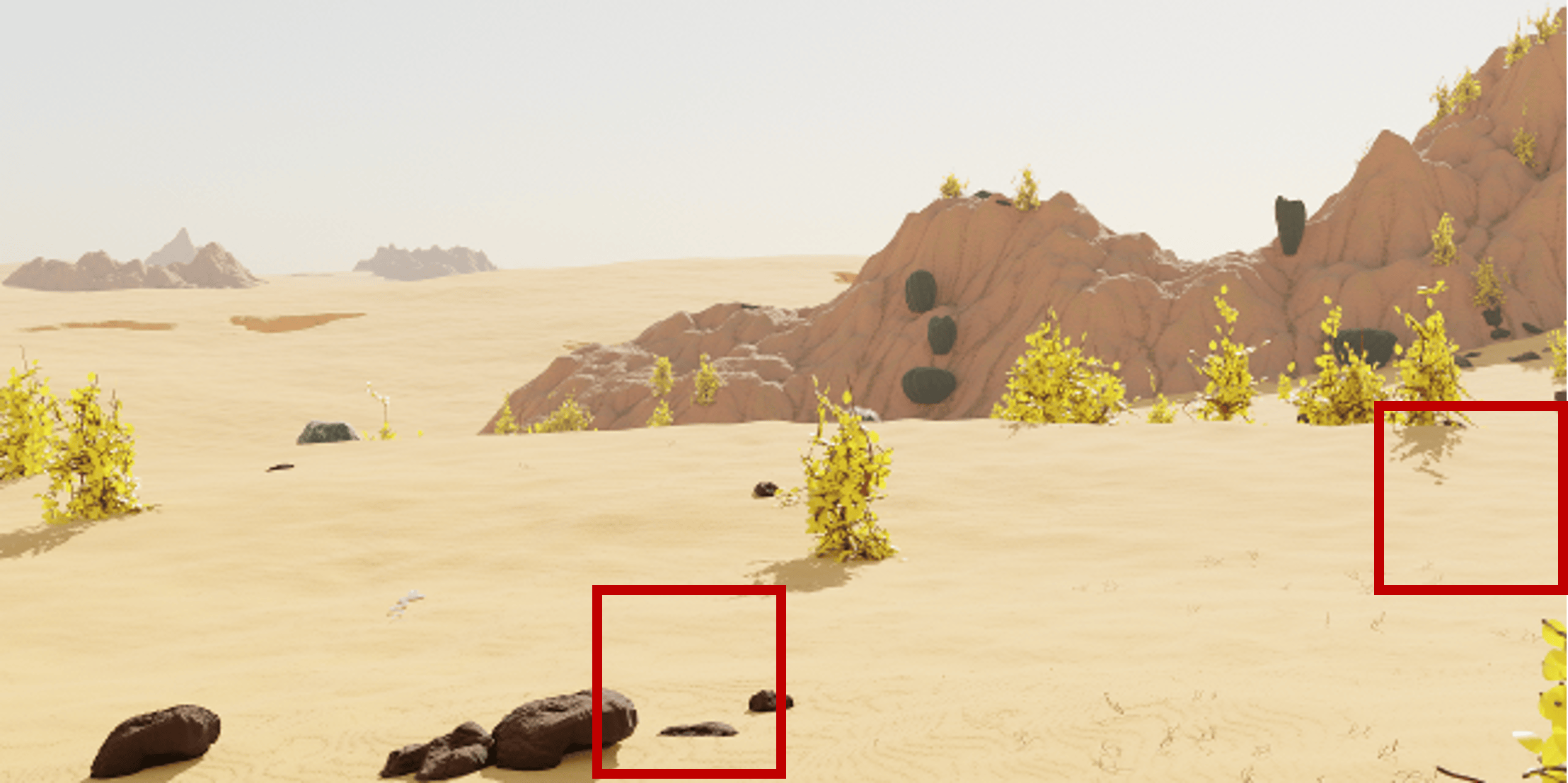}
  \\
  \raisebox{.4cm}{\makebox[0.01\textwidth]{\rotatebox{90}{\normalsize Depth}}}
  \includegraphics[width=\wratio\textwidth]{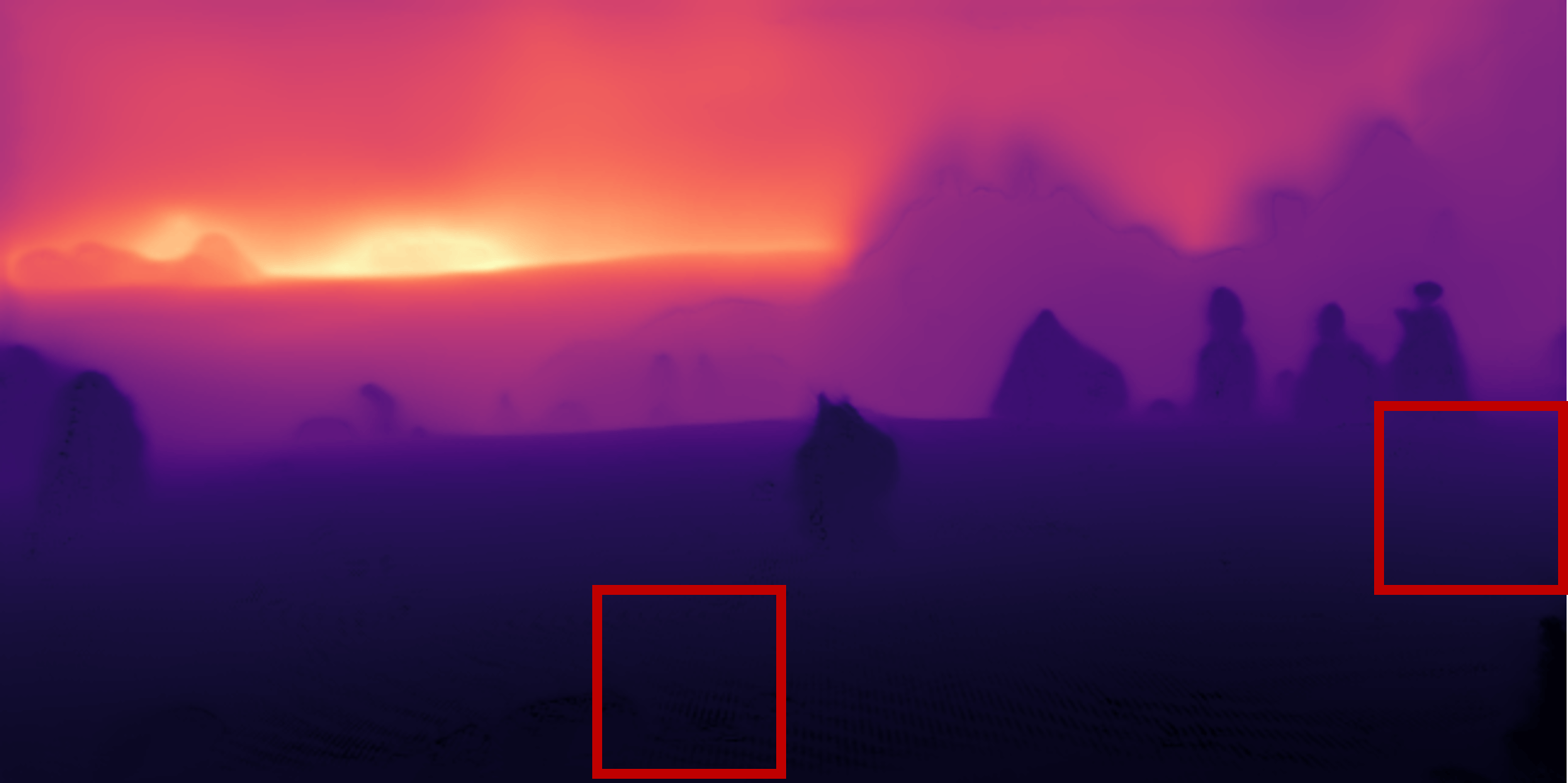}
  \includegraphics[width=\wratio\textwidth]{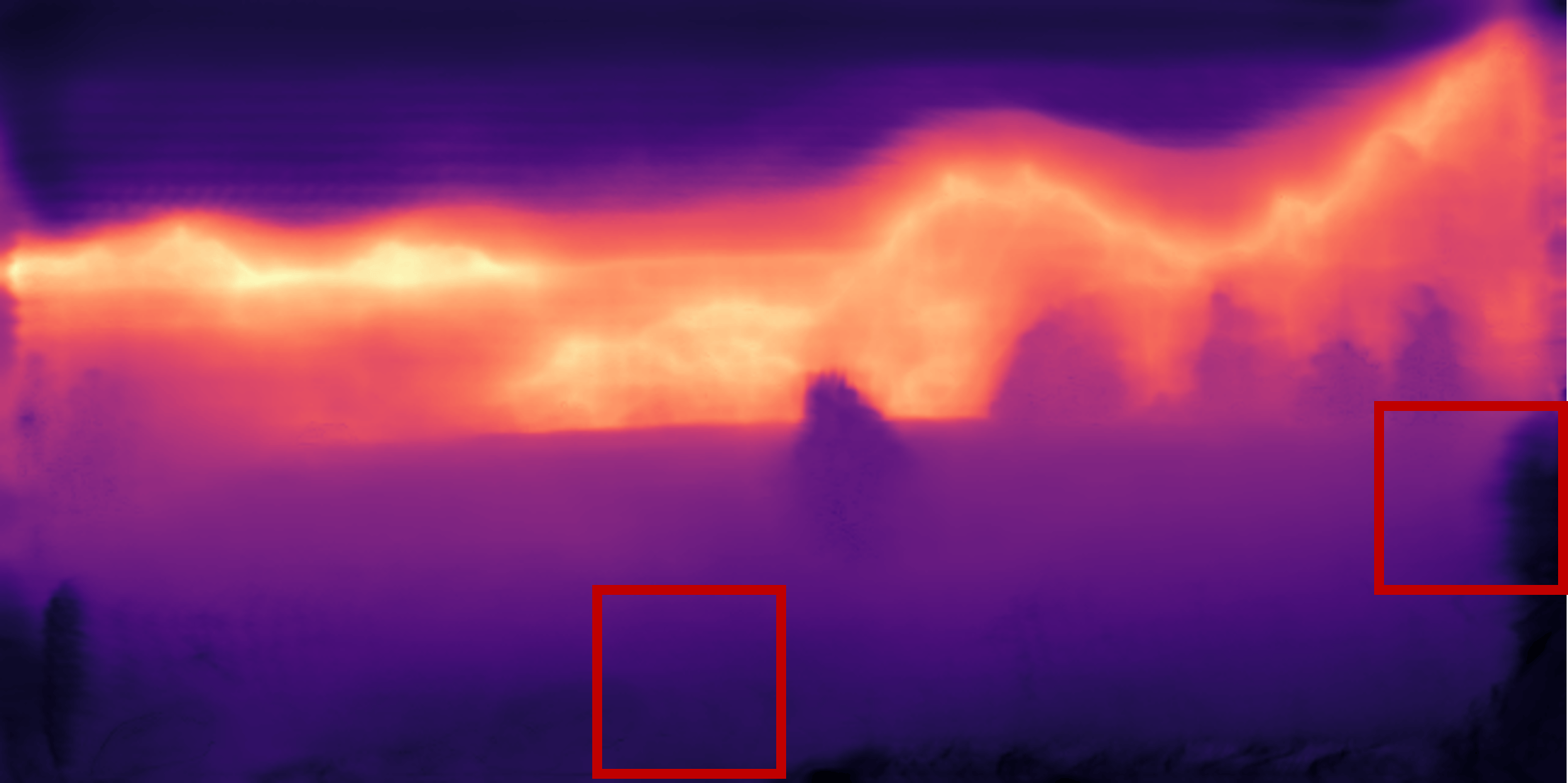}
  \includegraphics[width=\wratio\textwidth]{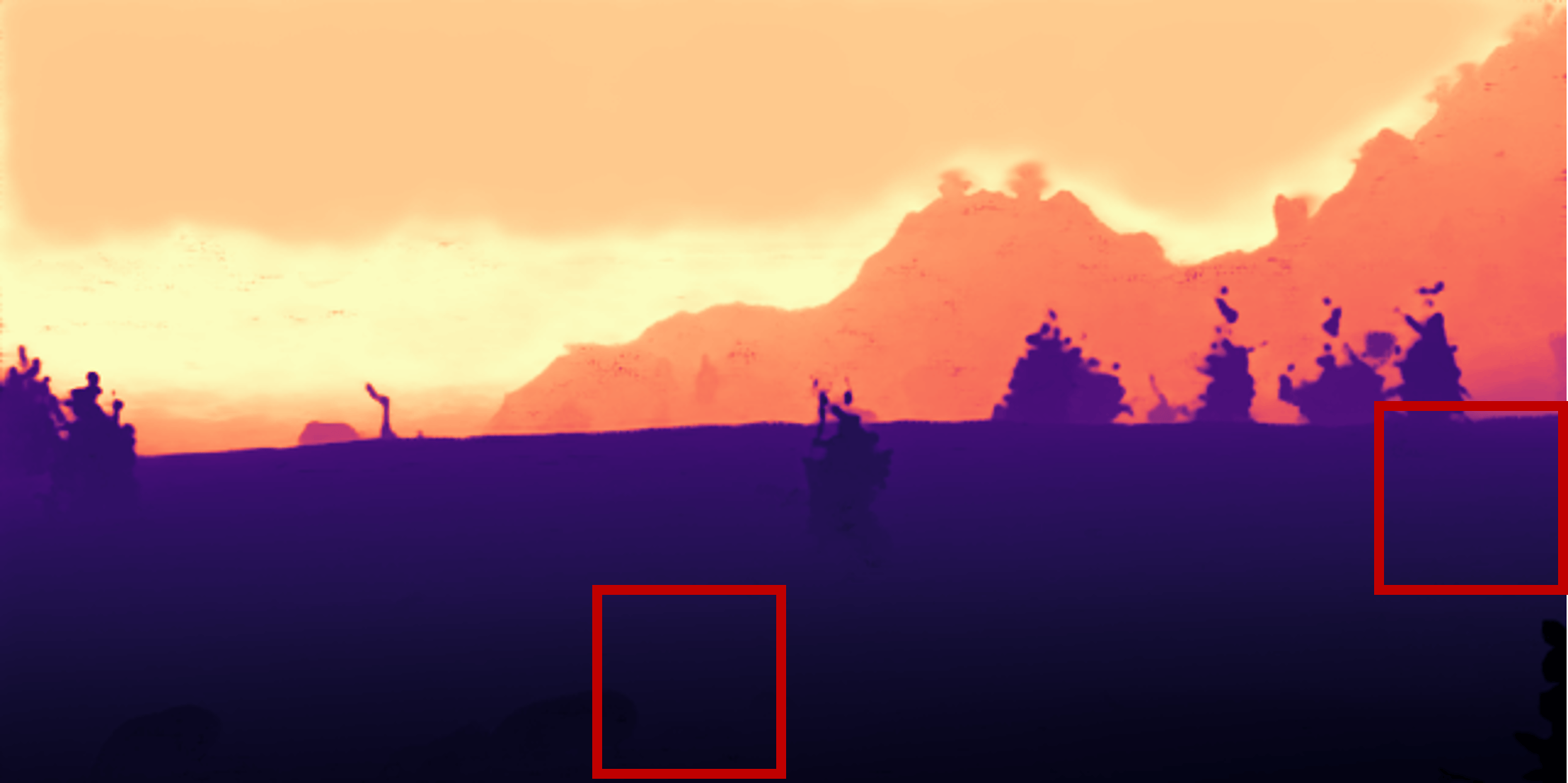}
  \includegraphics[width=\wratio\textwidth]{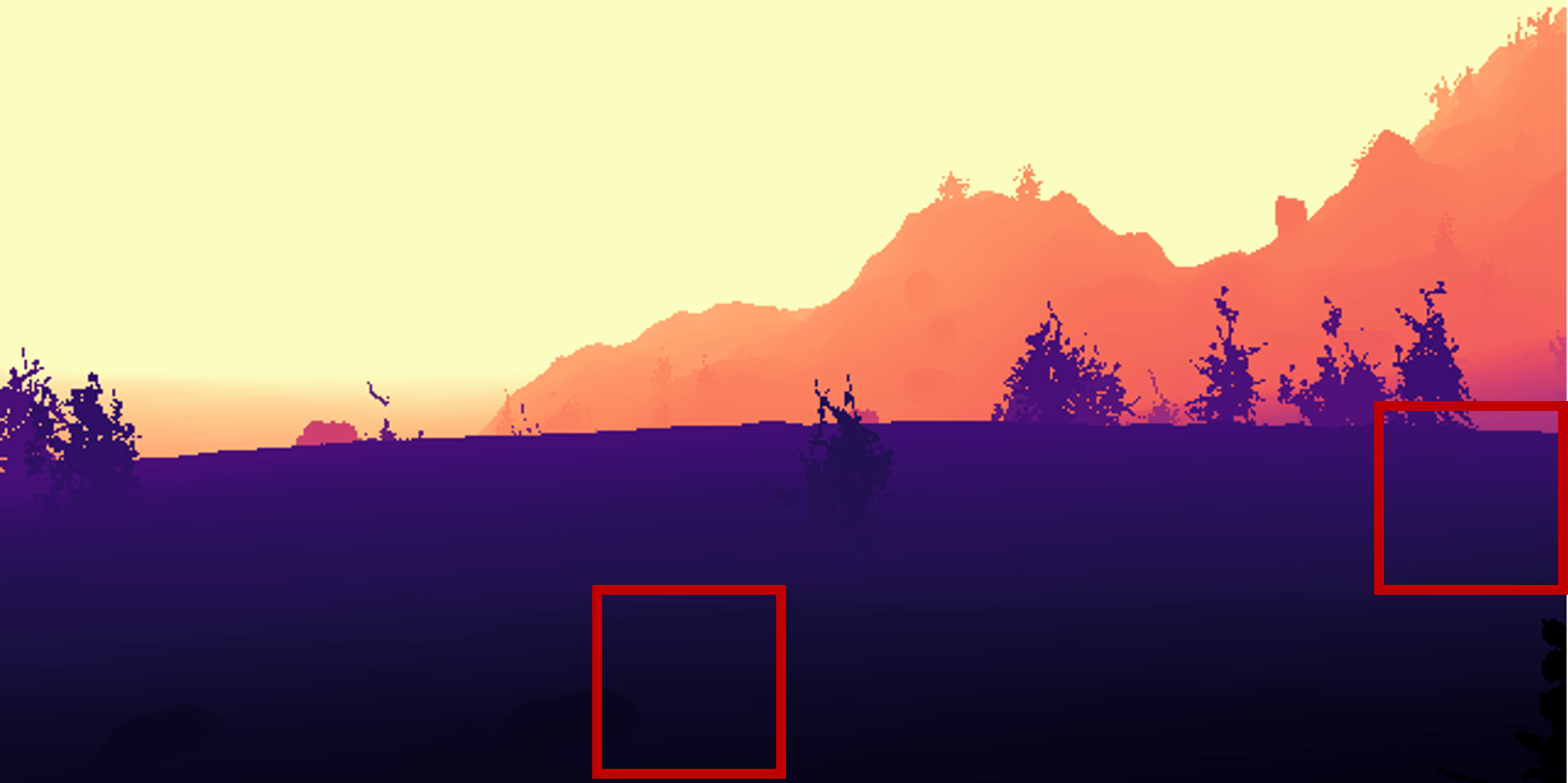}
  \\
  \caption{\textbf{Qualitative Results} on TartanAir and Infinigen. While the baselines can plausibly render the color of the novel views, they struggle to render accurate depth. \ours with a stereo-native architecture is capable of capturing both accurate depth and color. Methods marked with $^\dagger$ use ground-truth camera poses. Red squares highlight key regions. Best viewed zoomed in.}
  \label{fig:qualitative_results}
  \vspace{-10pt}
\end{figure*}

As shown in Table~\ref{tab:nvs_comparison}, \ours produces highly accurate novel-view geometry together with high rendering quality. Conversely, baseline methods consistently struggle with scene geometry across datasets. For pose-free models like VicaSplat and AnySplat, this limitation arises because their predicted representations exist in a normalized, non-metric space. This issue is further exacerbated by treating stereo pairs as independent frames, which leads to accumulated camera pose errors and $\delta_1$ scores below 10\% on Infinigen. Pose-aware architectures encounter similar geometric challenges. Although they use large 3D or depth foundation models as backbones, explicit depth supervision, and ground-truth poses, methods such as pixelSplat, MVSplat, and DepthSplat do not produce accurate geometry from the same stereo frames. Notably, their geometric performance remains remarkably poor even when they produce visually plausible color reconstructions. In contrast, \ours explicitly exploits the stereo structure and synthesizes novel views with highly accurate geometry. We analyze the contribution of each component in Sec.~\ref{sec:ablations}. Our approach consistently maintains accurate scene geometry across all tested environments, avoiding the severe degradation seen in baseline models (Fig ~\ref{fig:qualitative_results}). \ours also achieves the highest PSNR on all benchmarks, including the out-of-distribution real-world KITTI dataset.
\begin{table*}[h]
\centering
\footnotesize
\caption{\textbf{Novel view synthesis performance.} \ours achieves the most accurate novel-view geometry and the highest PSNR across all benchmarks. Methods marked with $^\dagger$ use ground truth camera poses. For methods marked with $^\ddagger$, depth is aligned to the ground truth with oracle median scaling.}
\label{tab:nvs_comparison}
\small
\setlength{\tabcolsep}{13.5pt}
\begin{tabular}{l|ccc|ccc}
\toprule
\multicolumn{1}{c|}{\multirow{2}{*}{Method}} & \multicolumn{3}{c|}{Novel View Geometry} & \multicolumn{3}{c}{Novel View Appearance} \\
\cmidrule(lr){2-4} \cmidrule(lr){5-7}
& AbsRel$\downarrow$ & $\delta_1$$\uparrow$ & L1$\downarrow$ & PSNR$\uparrow$ & SSIM$\uparrow$ & LPIPS$\downarrow$ \\
\midrule
\multicolumn{7}{l}{\textit{Infinigen}} \\
VicaSplat~\citep{li2025vicasplat} & 0.63 & 4.5 & 20.0 & 15.9 & 0.52 & 0.48 \\
AnySplat~\citep{jiang2025anysplat} & 0.53 & 0.5 & 15.9 & 25.2 & 0.88 & 0.18 \\
DA3~\citep{depthanything3}$^\ddagger$ & 0.41 & 42.2 & 9.5 & 21.6 & 0.73 & 0.33 \\
MVSplat~\citep{chen2024mvsplat}$^\dagger$ & 2.64 & 20.7 & 33.5 & 27.1 & 0.90 & 0.15 \\
pixelSplat~\citep{charatan23pixelsplat}$^\dagger$ & 0.47 & 40.1 & 10.8 & \underline{29.0} & \underline{0.95} & \textbf{0.07} \\
DepthSplat~\citep{xu2024depthsplat}$^\dagger$ & \underline{0.21} & \underline{63.0} & \underline{7.1} & 26.8 & 0.91 & 0.13 \\
Ours & \textbf{0.03} & \textbf{98.6} & \textbf{1.1} & \textbf{29.4} & \textbf{0.95} & \underline{0.12} \\
\midrule
\multicolumn{7}{l}{\textit{TartanV2}} \\
VicaSplat~\citep{li2025vicasplat} & 0.52 & 23.8 & 4.8 & 12.2 & 0.26 & 0.59 \\
AnySplat~\citep{jiang2025anysplat} & 0.22 & 61.9 & 2.8 & 18.6 & 0.66 & 0.26 \\
DA3~\citep{depthanything3}$^\ddagger$ & 0.47 & 40.8 & 3.6 & 18.3 & 0.59 & 0.32 \\
MVSplat~\citep{chen2024mvsplat}$^\dagger$ & 1.35 & 8.4 & 8.1 & 20.2 & 0.85 & 0.18 \\
pixelSplat~\citep{charatan23pixelsplat}$^\dagger$ & 0.36 & 70.2 & 2.7 & 21.8 & 0.91 & 0.14 \\
DepthSplat~\citep{xu2024depthsplat}$^\dagger$ & \underline{0.15} & \underline{78.7} & \underline{1.8} & \underline{25.8} & \underline{0.93} & \textbf{0.10} \\
Ours & \textbf{0.06} & \textbf{94.9} & \textbf{0.7} & \textbf{26.3} & \textbf{0.94} & \textbf{0.10} \\
\midrule
\multicolumn{7}{l}{\textit{KITTI}} \\
VicaSplat~\citep{li2025vicasplat} & 0.25 & 49.5 & 5.3 & 16.5 & 0.64 & 0.28 \\
AnySplat~\citep{jiang2025anysplat} & 0.18 & 71.2 & 3.5 & 18.1 & 0.70 & 0.35 \\
DA3~\citep{depthanything3}$^\ddagger$ & 0.32 & 44.6 & 5.2 & 17.0 & 0.58 & 0.33 \\
MVSplat~\citep{chen2024mvsplat}$^\dagger$ & 0.44 & 32.6 & 7.2 & 22.2 & \underline{0.92} & \textbf{0.09} \\
pixelSplat~\citep{charatan23pixelsplat}$^\dagger$ & 0.37 & 43.6 & 6.9 & \underline{22.3} & 0.91 & 0.11 \\
DepthSplat~\citep{xu2024depthsplat}$^\dagger$ & \underline{0.13} & \underline{79.8} & \underline{2.4} & 22.1 & 0.88 & 0.18 \\
Ours & \textbf{0.04} & \textbf{96.0} & \textbf{0.9} & \textbf{25.2} & \textbf{0.94} & \underline{0.10} \\
\bottomrule
\end{tabular}
\end{table*}
\subsection{Ablation Studies}
\label{sec:ablations}
\boldparagraph{Feature Selection.} The stereo encoder generates a feature pyramid $\{\mathbf{F}^{(s)}\}_{s=1}^{S}$ at four spatial resolutions per view. To evaluate our design, we first test a naive baseline that processes all scales independently without cross-view fusion, yielding the lowest performance. We then ablate our proposed cross-attention model by selectively discarding scales. This reveals that omitting the high-dimensional fine spatial features $\mathbf{F}_{s=0}$ degrades performance more than losing the coarsest semantic context $\mathbf{F}_{s=4}$. Ultimately, applying cross-attention across the entire multi-scale pyramid yields the best overall rendering quality, as shown in Table~\ref{tab:ablation_feature_interaction}.
\begin{table}[h]
    \centering
    \small
    
    \begin{minipage}[t]{0.48\textwidth}
        \centering
        \caption{Cross attention over multi-scale features across all input frames yields the best results.}
        \label{tab:ablation_feature_interaction}
        \setlength{\tabcolsep}{18pt}
        \begin{tabular}{lcc}
            \toprule
            Experiment & PSNR$\uparrow$ & L1$\downarrow$ \\
            \midrule
            No fusion & 22.8 & 1.4 \\
            Drop $\mathbf{F}_{s=0}$ & 22.9 & 1.1 \\
            Drop $\mathbf{F}_{s=4}$ & 23.1 & 1.1 \\
            Ours & \textbf{25.2} & \textbf{0.9} \\
            \bottomrule
        \end{tabular}
    \end{minipage}\hfill
    \begin{minipage}[t]{0.48\textwidth}
        \centering
        \caption{Adaptive aggregation enables high compression while improving rendering quality.}
        \label{tab:ablation_aggregation}
        \setlength{\tabcolsep}{6pt}
        \begin{tabular}{p{2.7cm} cc}
            \toprule
            Strategy & PSNR$\uparrow$ & Comp. Rate$\uparrow$ \\
            \midrule
            No Aggregation & 23.3 & 1.0 \\
            Voxelization & 23.2 & 1.1 \\
            Learnable Vox. & 23.3 & 1.1 \\
            Ours & \textbf{25.2} & \textbf{1.6} \\
            \bottomrule
        \end{tabular}
    \end{minipage}
    \vspace{-10pt}
\end{table}

\begin{figure}[h]
    \centering
    \begin{minipage}[t]{0.48\textwidth}
        \centering
        \includegraphics[width=\linewidth]{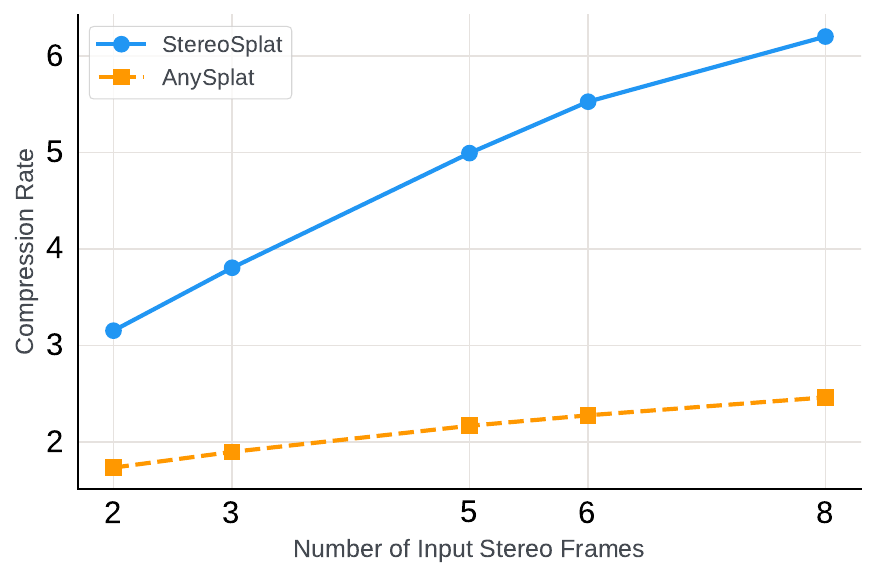}
        \vspace{-18pt}
        \caption{Our adaptive voxelization enables \ours to achieve superior compression which scales
        as more stereo frames are provided.}
        \label{fig:vis_compression}
    \end{minipage}\hfill
    \begin{minipage}[t]{0.48\textwidth}
        \centering
        \includegraphics[width=\linewidth]{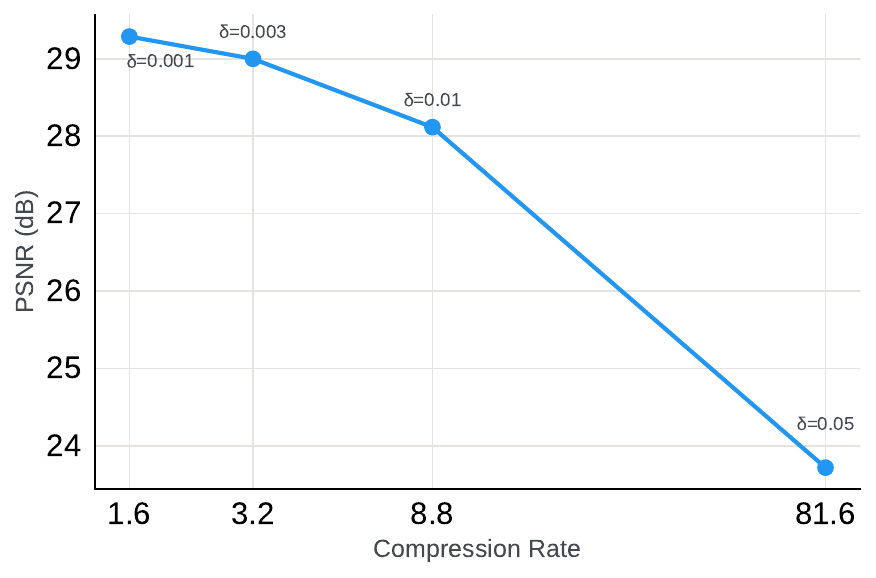}
        \vspace{-18pt}
        \caption{\ours offers a flexible balance between compression rate and rendering quality by tuning the initial voxel size $\delta$.}
        \label{fig:viz_psnr_scaling}
    \end{minipage}
\end{figure}

\boldparagraph{Aggregation Mechanism.} We evaluate several strategies for aggregating the primitives predicted at each pixel to improve spatial efficiency. The baseline retains all generated primitives without any reduction. We also compare against standard voxelization and the learnable variant introduced by \citet{jiang2025anysplat}. We then evaluate our proposed adaptive method, which dynamically scales voxel resolution using depth and appearance features. As demonstrated in Table~\ref{tab:ablation_aggregation}, our approach achieves the highest rendering performance. This improvement arises from merging primitives in distant regions and clustering them using both geometric and visual cues. Furthermore, this strategy significantly increases the overall compression rate as visualized in Figure~\ref{fig:vis_compression}.

\boldparagraph{Compression Rate.} Our approach provides a flexible trade-off between rendering quality and compression rate, defined as the ratio of retained primitives to the initial dense predictions. By adjusting the initial voxel size, we can significantly reduce the primitive count while incurring only a marginal decrease in reconstruction accuracy (Fig.~\ref{fig:viz_psnr_scaling}). AnySplat~\citep{jiang2025anysplat} serves as the only baseline that explicitly compresses the 3D representation. We compare our method against this baseline on the Infinigen dataset. The camera motion between frames in these sequences remains relatively small across multiple input views. This minimal movement creates high visual redundancy, making the dataset an ideal environment for evaluating the effectiveness of compression mechanisms. Our approach demonstrates superior compression efficiency and scales effectively with the number of views, as shown in Fig.~\ref{fig:vis_compression}. This advantage stems directly from our adaptive voxelization strategy, which preserves critical structural details while discarding redundant primitives.

\paragraph{Scaling to Longer Videos.} Our approach scales effectively to long video sequences. Thanks to the high geometric accuracy and pose-aware design of our framework, we perform inference independently across temporal chunks, concatenate the predictions, and refine the full aggregated scene with a small number of optimization iterations. As shown in Fig.~\ref{fig:teaser}, this approach successfully consolidates over 600 input frames into a globally consistent representation.

\paragraph{Limitations and Future Work.} Although \ours demonstrates strong NVS performance, it inherently relies on calibrated stereo pairs, precluding its use with uncalibrated or monocular captures. Furthermore, because our method assumes a static environment during scene aggregation, reconstruction quality degrades when processing highly dynamic videos. In addition, its end-to-end reconstruction speed is bounded by the stereo backbone, although our aggregation yields the lowest memory footprint and the fastest rendering among the compared methods (see supplementary material). An exciting direction for future work is to relax the calibration requirement and extend our framework to dynamic scenes.

\section{Conclusion}
We presented \ours, the first feed-forward 3D Gaussian Splatting framework designed for stereo videos. \ours exploits the metric scale of calibrated stereo rigs and preserves it throughout the pipeline, overcoming the scale ambiguity of previous methods to predict accurate 3D Gaussians in metric space. Furthermore, our scene aggregation mechanism efficiently compresses per-pixel primitives, while pre-training on our large-scale synthetic \oursdataset dataset improves generalization to real-world data. Extensive evaluations demonstrate that \ours achieves state-of-the-art geometric accuracy and superior novel view synthesis across diverse synthetic and real-world scenes.

\section*{Acknowledgments}

We thank Jaap Suter, Daniel Scharstein, Conor Oberlander, and David Fuhrer at Meta for creating the XRStereo dataset used in this work. Their work on the rendering pipeline, art design, and the variety of stereo rig configurations made the large-scale stereo video pre-training in this paper possible.

\bibliographystyle{plainnat}
\bibliography{main}

\newpage
\appendix

\section{Supplementary Material}

\subsection{Implementation Details}
We train and evaluate our model on $320 \times 640$ resolution images with three input views per sample. The model is optimized using \textsc{AdamW} with a total effective batch size of $40$. The learning rate follows a 10-epoch schedule, starting with a one-epoch linear warm-up to $\eta_{\max}=10^{-3}$, followed by a cosine decay to $\eta_{\min}=10^{-5}$ over the remaining budget. The total training objective is defined as:
\begin{equation}
\mathcal{L} = 0.8\,\mathcal{L}_{\ell_1}^{\text{rgb}} + 0.2\,\mathcal{L}_{\text{ssim}}^{\text{rgb}} + 0.05\,\mathcal{L}_{\text{lpips}}^{\text{rgb}} + 0.5\,\mathcal{L}_{\ell_1}^{\text{depth}},
\end{equation}
where the depth term $\mathcal{L}_{\ell_1}^{\text{depth}}$ is masked to include only pixels with valid ground-truth depth within the range $[0, 80]$\,m. The loss weights are determined empirically based on validation performance. The framework is implemented in PyTorch and uses the original implementation from FoundationStereo as backbone, and the total training time is approximately $240$ GPU-hours on NVIDIA H100.

\FloatBarrier
\subsection{Additional Quantitative Evaluation} To further evaluate the robustness of \ours, we provide additional experiments across varying temporal window sizes $N \in \{2, 3, 4\}$, corresponding to total image counts of $4, 6,$ and $8$, respectively. This evaluation is conducted in a progressive manner, where for a window of size $N$, we utilize $N-1$ intermediate frames for evaluation, which effectively increases the task difficulty by expanding the temporal baseline and increasing the number of 3DGS primitives. For improved readability, we categorize our results into geometric accuracy (Table~\ref{supmat_tab:geometry_comparison}) and appearance quality (Table~\ref{supmat_tab:nvs_comparison}). In both tables, we include a baseline comparison of the model's performance when trained without \oursdataset to quantify the impact of our proposed data. While this variant remains competitive on the in-distribution Infinigen and TartanV2 benchmarks, the gap widens substantially on KITTI, indicating that \oursdataset is particularly beneficial for generalization to unseen real-world data. While the reported experiments are constrained by standard evaluation protocols, our model is inherently scalable and supports an arbitrary number of input views, bounded only by the available GPU memory.
\begin{table*}[ht]
\centering
\footnotesize
\caption{\textbf{Novel view geometry.} \ours achieves the most accurate novel-view geometry across all benchmarks. Methods marked with $^\dagger$ use ground truth camera poses.}
\label{supmat_tab:geometry_comparison}
\resizebox{\textwidth}{!}{
\begin{tabular}{l|ccc|ccc|ccc}
\toprule
\multicolumn{1}{c|}{\multirow{2}{*}{Method}} & \multicolumn{3}{c|}{2 Stereo Frames} & \multicolumn{3}{c|}{3 Stereo Frames} & \multicolumn{3}{c}{4 Stereo Frames} \\
\cmidrule(lr){2-4} \cmidrule(lr){5-7} \cmidrule(lr){8-10}
& AbsRel$\downarrow$ & $\delta_1$$\uparrow$ & L1$\downarrow$ & AbsRel$\downarrow$ & $\delta_1$$\uparrow$ & L1$\downarrow$ & AbsRel$\downarrow$ & $\delta_1$$\uparrow$ & L1$\downarrow$ \\
\midrule
\multicolumn{10}{l}{\textit{Infinigen}} \\
VicaSplat & 0.63 & 4.5 & 20.0 & 0.60 & 8.4 & 18.9 & 0.72 & 2.7 & 23.1 \\
AnySplat & 0.53 & 0.5 & 15.9 & 0.50 & 1.4 & 15.0 & 0.48 & 1.7 & 14.7 \\
MVSplat$^\dagger$ & 2.64 & 20.7 & 33.5 & 2.54 & 20.5 & 32.4 & 2.63 & 20.0 & 33.6 \\
pixelSplat$^\dagger$ & 0.47 & 40.1 & 10.8 & 0.48 & 39.7 & 10.8 & 0.48 & 39.7 & 10.8 \\
DepthSplat$^\dagger$ & 0.21 & 63.0 & 7.1 & 0.26 & 52.5 & 10.2 & 0.29 & 49.6 & 10.6 \\
Ours w/o \oursdataset & \underline{0.04} & \underline{97.5} & \textbf{1.1} & \underline{0.04} & \underline{97.5} & \underline{1.1} & \underline{0.04} & \underline{97.7} & \underline{1.1} \\
Ours & \textbf{0.03} & \textbf{98.6} & \textbf{1.1} & \textbf{0.03} & \textbf{98.6} & \textbf{1.0} & \textbf{0.03} & \textbf{98.7} & \textbf{1.0} \\
\midrule
\multicolumn{10}{l}{\textit{TartanV2}} \\
VicaSplat & 0.52 & 23.8 & 4.8 & 0.55 & 19.4 & 5.1 & 0.55 & 28.7 & 4.5 \\
AnySplat & 0.22 & 61.9 & 2.8 & 0.21 & 66.4 & 2.7 & 0.20 & 67.0 & 2.6 \\
MVSplat$^\dagger$ & 1.35 & 8.4 & 8.1 & 1.42 & 10.2 & 8.5 & 1.54 & 10.7 & 8.7 \\
pixelSplat$^\dagger$ & 0.36 & 70.2 & 2.7 & 0.57 & 63.5 & 3.1 & 0.66 & 60.8 & 3.3 \\
DepthSplat$^\dagger$ & 0.15 & 78.7 & 1.8 & 0.15 & 80.5 & 1.8 & 0.13 & 82.7 & 1.8 \\
Ours w/o \oursdataset & \textbf{0.06} & \underline{94.8} & \textbf{0.7} & \textbf{0.08} & \underline{94.4} & \textbf{0.7} & \textbf{0.06} & \textbf{94.8} & \textbf{0.7} \\
Ours & \textbf{0.06} & \textbf{94.9} & \textbf{0.7} & \textbf{0.08} & \textbf{94.5} & \textbf{0.7} & \textbf{0.06} & \textbf{94.8} & \textbf{0.7} \\
\midrule
\multicolumn{10}{l}{\textit{KITTI}} \\
VicaSplat & 0.25 & 49.5 & 5.3 & 0.35 & 22.6 & 7.5 & 0.34 & 43.4 & 5.4 \\
AnySplat & 0.18 & 71.2 & 3.5 & 0.13 & 82.4 & 2.8 & 0.14 & 79.1 & 3.0 \\
MVSplat$^\dagger$ & 0.44 & 32.6 & 7.2 & 0.37 & 38.0 & 6.4 & 0.37 & 37.3 & 6.4 \\
pixelSplat$^\dagger$ & 0.37 & 43.6 & 6.9 & 0.44 & 35.6 & 7.9 & 0.47 & 33.0 & 8.4 \\
DepthSplat$^\dagger$ & 0.13 & 79.8 & 2.4 & 0.15 & 76.1 & 2.6 & 0.20 & 64.1 & 3.3 \\
Ours w/o \oursdataset & \underline{0.06} & \underline{94.7} & \underline{1.1} & \underline{0.07} & \underline{91.7} & \underline{1.2} & \underline{0.07} & \underline{91.6} & \underline{1.2} \\
Ours & \textbf{0.04} & \textbf{96.0} & \textbf{0.9} & \textbf{0.05} & \textbf{95.7} & \textbf{1.0} & \textbf{0.04} & \textbf{95.7} & \textbf{1.0} \\
\bottomrule
\end{tabular}
}
\end{table*}
\begin{table*}[ht]
\centering
\footnotesize
\caption{\textbf{Novel view appearance.} Methods marked with $^\dagger$ use ground truth camera poses.}
\label{supmat_tab:nvs_comparison}
\resizebox{\textwidth}{!}{
\begin{tabular}{l|ccc|ccc|ccc}
\toprule
\multicolumn{1}{c|}{\multirow{2}{*}{Method}} & \multicolumn{3}{c|}{2 Stereo Frames} & \multicolumn{3}{c|}{3 Stereo Frames} & \multicolumn{3}{c}{4 Stereo Frames} \\
\cmidrule(lr){2-4} \cmidrule(lr){5-7} \cmidrule(lr){8-10}
& PSNR$\uparrow$ & SSIM$\uparrow$ & LPIPS$\downarrow$ & PSNR$\uparrow$ & SSIM$\uparrow$ & LPIPS$\downarrow$ & PSNR$\uparrow$ & SSIM$\uparrow$ & LPIPS$\downarrow$ \\
\midrule
\multicolumn{10}{l}{\textit{Infinigen}} \\
VicaSplat & 15.9 & 0.52 & 0.48 & 16.4 & 0.52 & 0.50 & 17.1 & 0.53 & 0.48 \\
AnySplat & 25.2 & 0.88 & 0.18 & 24.7 & 0.86 & 0.19 & 25.4 & 0.88 & 0.19 \\
MVSplat$^\dagger$ & 27.1 & 0.90 & 0.15 & 26.1 & 0.87 & 0.17 & 25.5 & 0.86 & 0.19 \\
pixelSplat$^\dagger$ & \underline{29.0} & \textbf{0.95} & \textbf{0.07} & 28.2 & \underline{0.94} & \textbf{0.09} & 28.1 & \underline{0.93} & \textbf{0.09} \\
DepthSplat$^\dagger$ & 26.8 & 0.91 & 0.13 & 26.9 & 0.91 & 0.14 & 26.7 & 0.91 & 0.15 \\
Ours w/o \oursdataset & 28.6 & 0.94 & 0.18 & \underline{28.4} & \underline{0.94} & 0.18 & \underline{28.2} & \underline{0.93} & 0.18 \\
Ours & \textbf{29.4} & \textbf{0.95} & \underline{0.12} & \textbf{29.0} & \textbf{0.95} & \underline{0.11} & \textbf{28.7} & \textbf{0.95} & \underline{0.12} \\
\midrule
\multicolumn{10}{l}{\textit{TartanV2}} \\
VicaSplat & 12.2 & 0.26 & 0.59 & 12.1 & 0.24 & 0.60 & 12.2 & 0.24 & 0.60 \\
AnySplat & 18.6 & 0.66 & 0.26 & 17.5 & 0.58 & 0.32 & 17.2 & 0.56 & 0.35 \\
MVSplat$^\dagger$ & 20.2 & 0.85 & 0.18 & 17.7 & 0.72 & 0.31 & 17.0 & 0.67 & 0.35 \\
pixelSplat$^\dagger$ & 21.8 & 0.91 & 0.14 & 19.2 & 0.80 & 0.26 & 18.4 & 0.76 & 0.30 \\
DepthSplat$^\dagger$ & \underline{25.8} & \underline{0.93} & \textbf{0.10} & \underline{25.0} & 0.91 & \textbf{0.12} & \textbf{25.3} & \underline{0.92} & \textbf{0.12} \\
Ours w/o \oursdataset & \textbf{26.3} & \textbf{0.94} & \underline{0.11} & \underline{25.0} & \underline{0.92} & \underline{0.13} & 24.9 & \underline{0.92} & \textbf{0.12} \\
Ours & \textbf{26.3} & \textbf{0.94} & \textbf{0.10} & \textbf{25.2} & \textbf{0.93} & \underline{0.13} & \underline{25.0} & \textbf{0.93} & \textbf{0.12} \\
\midrule
\multicolumn{10}{l}{\textit{KITTI}} \\
VicaSplat & 16.5 & 0.64 & 0.28 & 13.5 & 0.42 & 0.42 & 14.2 & 0.43 & 0.42 \\
AnySplat & 18.1 & 0.70 & 0.35 & 17.1 & 0.63 & 0.40 & 16.6 & 0.60 & 0.44 \\
MVSplat$^\dagger$ & 22.2 & 0.92 & \textbf{0.09} & 18.5 & 0.82 & \underline{0.18} & 17.4 & 0.77 & 0.23 \\
pixelSplat$^\dagger$ & 22.3 & 0.91 & 0.11 & 19.4 & 0.80 & 0.21 & 18.5 & 0.75 & 0.26 \\
DepthSplat$^\dagger$ & 22.1 & 0.88 & 0.18 & 21.3 & 0.84 & 0.21 & 19.4 & 0.78 & 0.27 \\
Ours w/o \oursdataset & \underline{24.4} & \underline{0.93} & 0.12 & \underline{22.4} & \underline{0.89} & \underline{0.18} & \underline{21.3} & \underline{0.88} & \underline{0.20} \\
Ours & \textbf{25.2} & \textbf{0.94} & \underline{0.10} & \textbf{23.9} & \textbf{0.92} & \textbf{0.13} & \textbf{22.9} & \textbf{0.90} & \textbf{0.14} \\
\bottomrule
\end{tabular}
}
\end{table*}
\subsection{Additional Qualitative Comparisons} 
\begin{figure*}[h]
  \centering
  \newcommand{\wratio}{0.24} 

  \makebox[0.01\textwidth][c]{}
  \makebox[\wratio\textwidth]{\normalsize AnySplat}
  \makebox[\wratio\textwidth]{\normalsize DepthSplat $\dagger$}
  \makebox[\wratio\textwidth]{\normalsize \ours (Ours)}
  \makebox[\wratio\textwidth]{\normalsize Ground Truth}
  \\
  \raisebox{.4cm}{\makebox[0.01\textwidth]{\rotatebox{90}{\normalsize Color}}}
  \includegraphics[width=\wratio\textwidth]{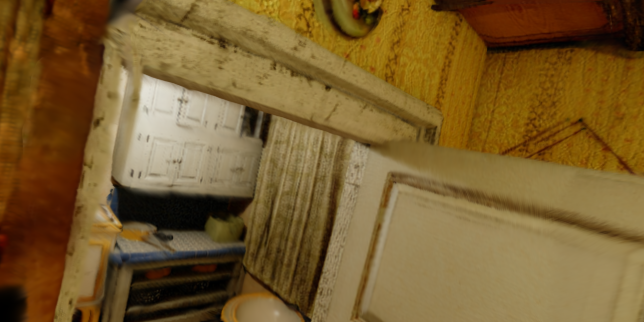}
  \includegraphics[width=\wratio\textwidth]{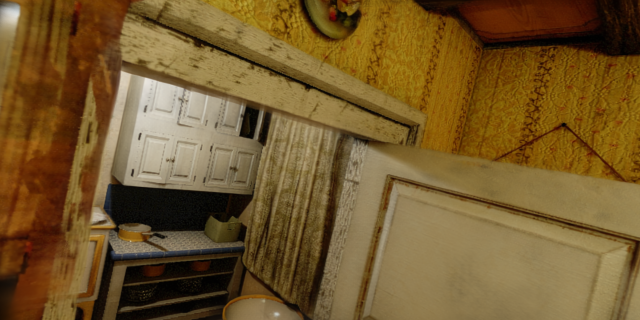}
  \includegraphics[width=\wratio\textwidth]{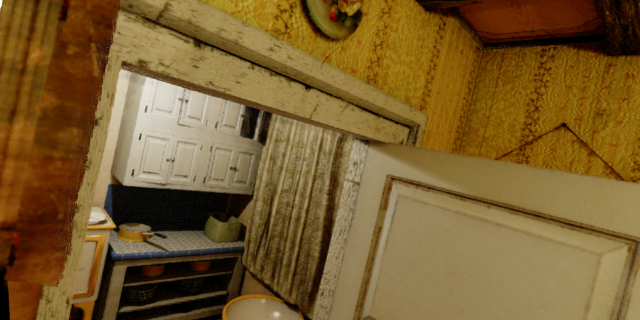}
  \includegraphics[width=\wratio\textwidth]{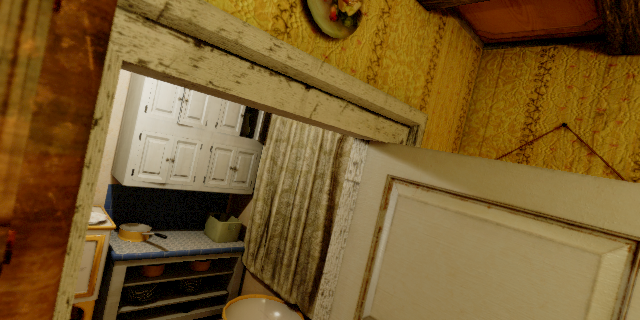}
  \\
  \raisebox{.4cm}{\makebox[0.01\textwidth]{\rotatebox{90}{\normalsize Depth}}}
  \includegraphics[width=\wratio\textwidth]{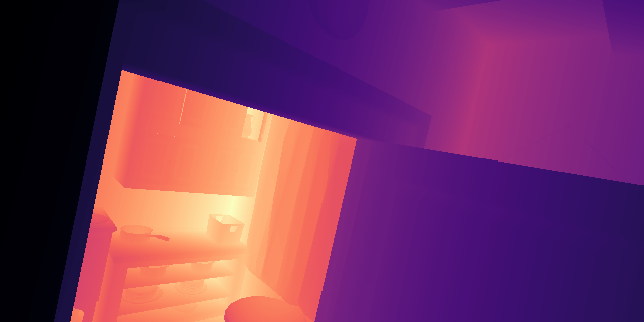}
  \includegraphics[width=\wratio\textwidth]{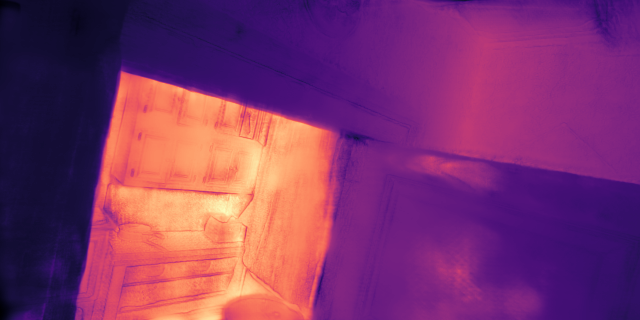}
  \includegraphics[width=\wratio\textwidth]{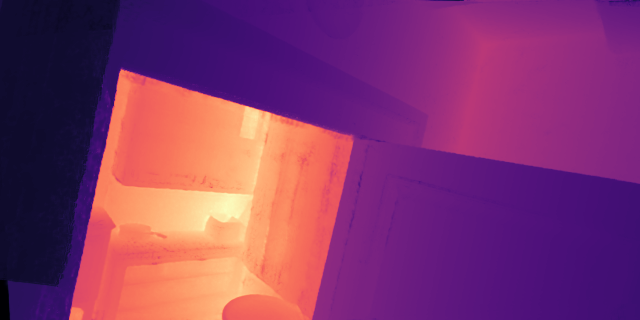}
  \includegraphics[width=\wratio\textwidth]{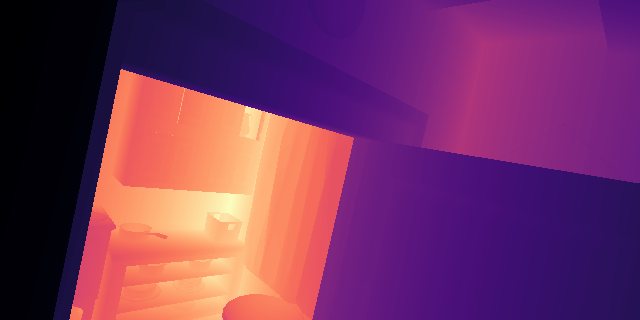}  
  \\  
  \raisebox{.4cm}{\makebox[0.01\textwidth]{\rotatebox{90}{\normalsize Color}}}
  \includegraphics[width=\wratio\textwidth]{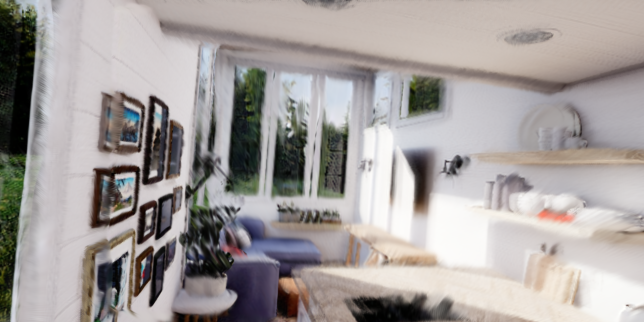}
  \includegraphics[width=\wratio\textwidth]{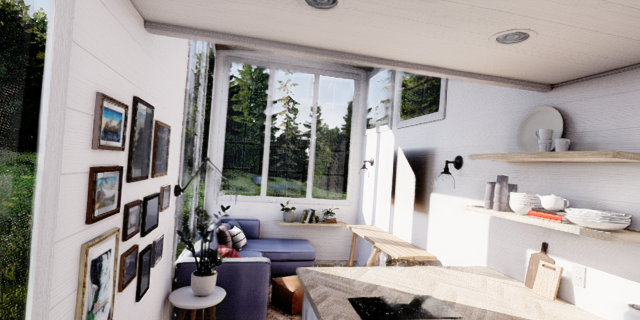}
  \includegraphics[width=\wratio\textwidth]{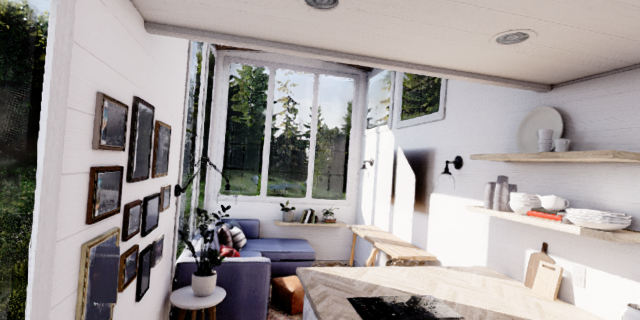}
  \includegraphics[width=\wratio\textwidth]{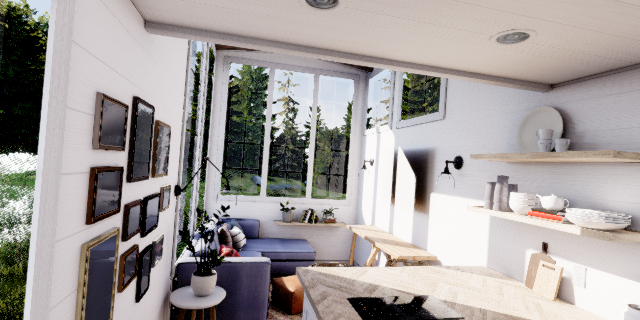}
  \\
  \raisebox{.4cm}{\makebox[0.01\textwidth]{\rotatebox{90}{\normalsize Depth}}}
  \includegraphics[width=\wratio\textwidth]{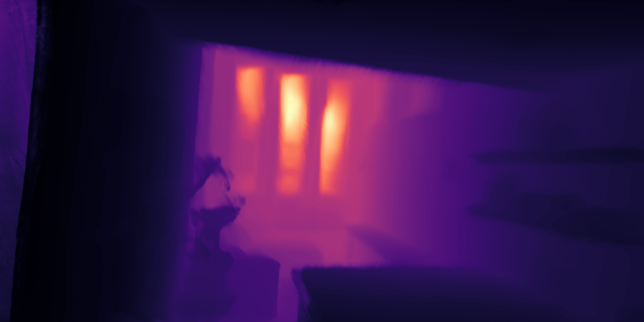}
  \includegraphics[width=\wratio\textwidth]{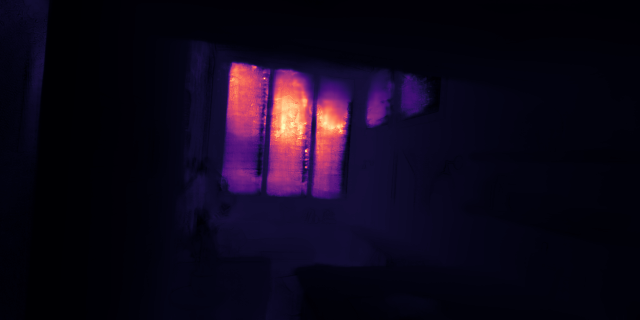}
  \includegraphics[width=\wratio\textwidth]{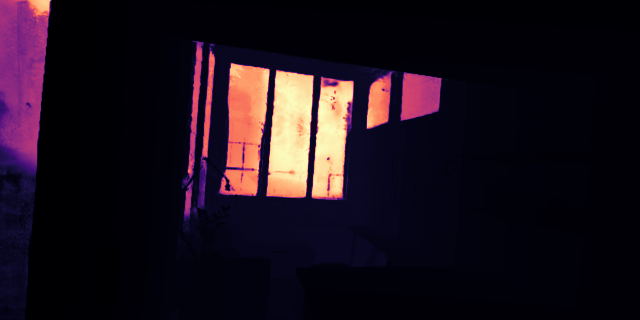}
  \includegraphics[width=\wratio\textwidth]{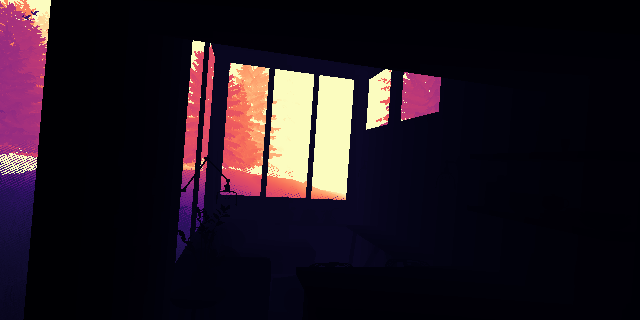}  
  \\
  \raisebox{.4cm}{\makebox[0.01\textwidth]{\rotatebox{90}{\normalsize Color}}}
  \includegraphics[width=\wratio\textwidth]{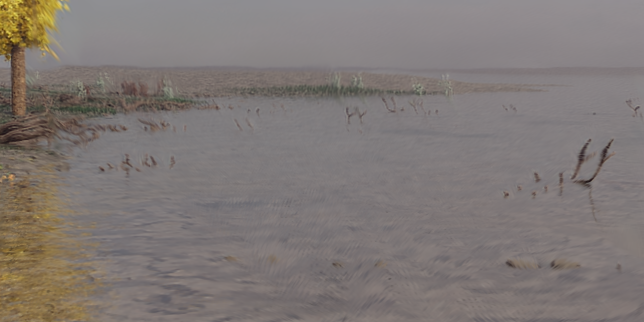}
  \includegraphics[width=\wratio\textwidth]{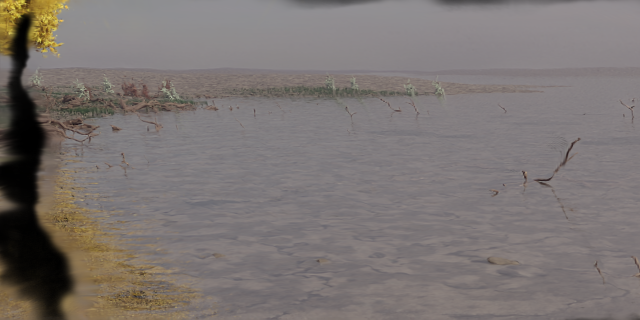}
  \includegraphics[width=\wratio\textwidth]{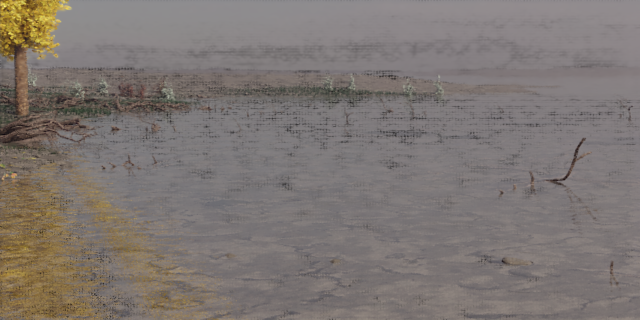}
  \includegraphics[width=\wratio\textwidth]{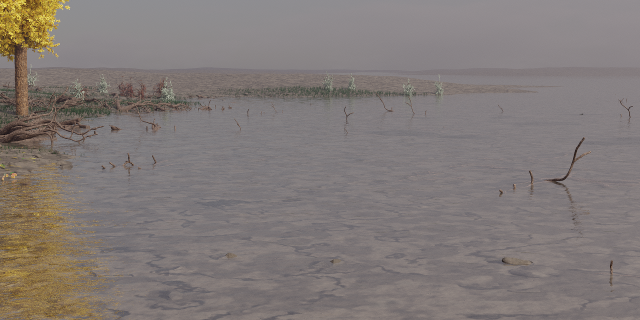}
  \\
  \raisebox{.4cm}{\makebox[0.01\textwidth]{\rotatebox{90}{\normalsize Depth}}}
  \includegraphics[width=\wratio\textwidth]{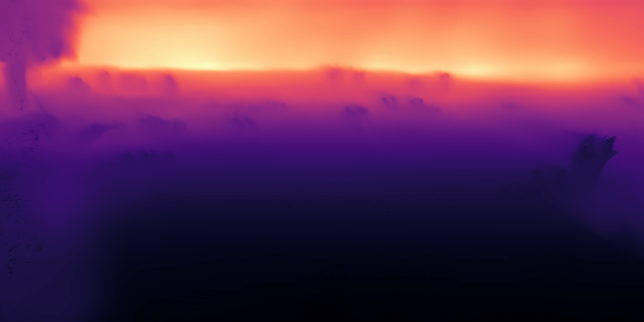}
  \includegraphics[width=\wratio\textwidth]{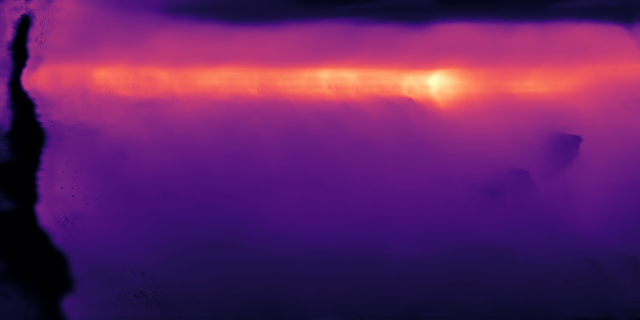}
  \includegraphics[width=\wratio\textwidth]{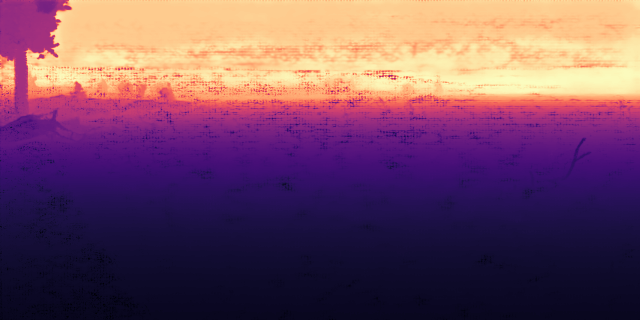}
  \includegraphics[width=\wratio\textwidth]{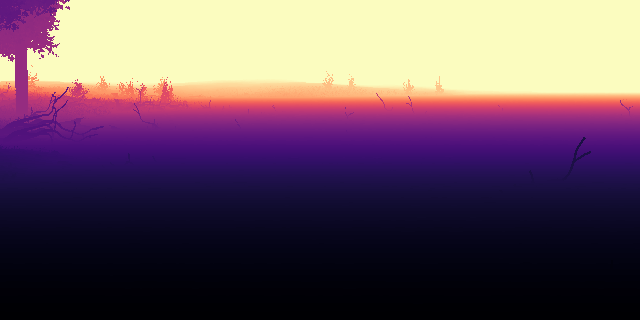}
  \\
  \caption{\textbf{Qualitative Results} on TartanAir and Infinigen. While the baselines can plausibly render the color of the novel views, they struggle to render accurate depth. \ours with a stereo-native architecture is capable of capturing both accurate color and depth. Methods marked with $^\dagger$ use ground-truth camera poses. Best viewed zoomed in.}
  \label{supmat_fig:qualitative_results}
\end{figure*}

We provide extended qualitative comparisons to further evaluate the visual fidelity of our method. For these visualizations, we focus on the most competitive baselines, AnySplat and DepthSplat. We omit VicaSplat, PixelSplat, and MVSplat from this comparison as these methods are natively restricted to a $256 \times 256$ resolution, which precludes a fair high-resolution comparison with our $320 \times 640$ results. Furthermore, these models consistently underperform relative to more recent state-of-the-art approaches. As illustrated in Figure~\ref{supmat_fig:qualitative_results}, AnySplat exhibits significant artifacts, particularly in rows 3 and 4, stemming from inaccuracies in camera pose estimation. While DepthSplat produces plausible color renderings, its underlying geometry is often inconsistent. This geometric inaccuracy manifests as photometric artifacts. For instance, the depth errors in row 5 directly degrade the clarity and structural integrity of the synthesized color view. In contrast, \ours maintains both sharp geometric boundaries and high color fidelity.
\FloatBarrier
\subsection{\oursdataset}
\begin{figure*}[h]
  \centering
  \newcommand{\wratio}{0.32}

  \raisebox{.4cm}{\makebox[0.01\textwidth]{\rotatebox{90}{\normalsize Color}}}
  \includegraphics[width=\wratio\textwidth]{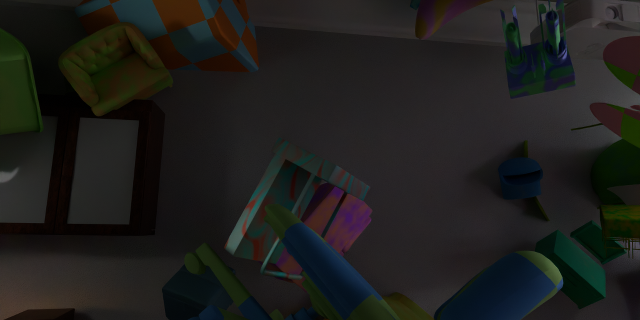}
  \includegraphics[width=\wratio\textwidth]{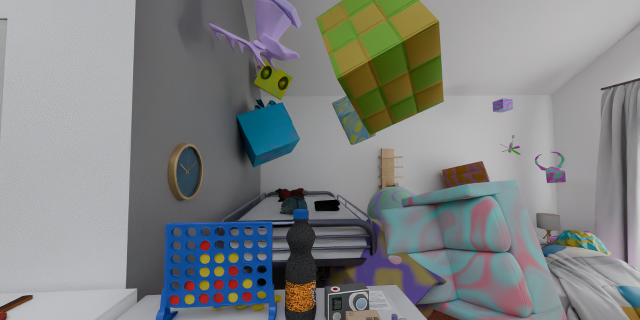}
  \includegraphics[width=\wratio\textwidth]{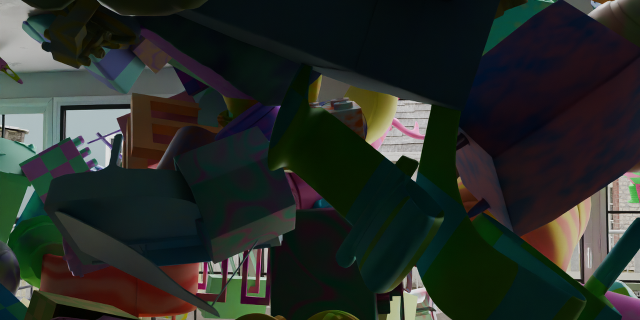}
  \\
  \raisebox{.4cm}{\makebox[0.01\textwidth]{\rotatebox{90}{\normalsize Depth}}}
  \includegraphics[width=\wratio\textwidth]{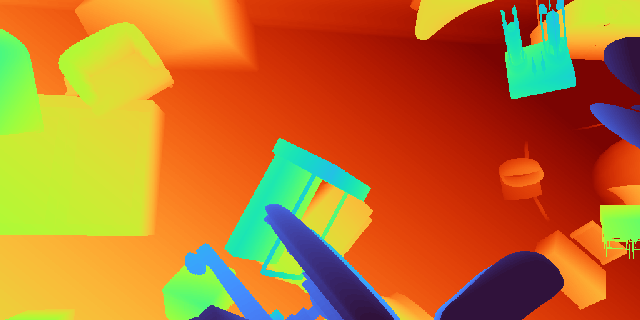}
  \includegraphics[width=\wratio\textwidth]{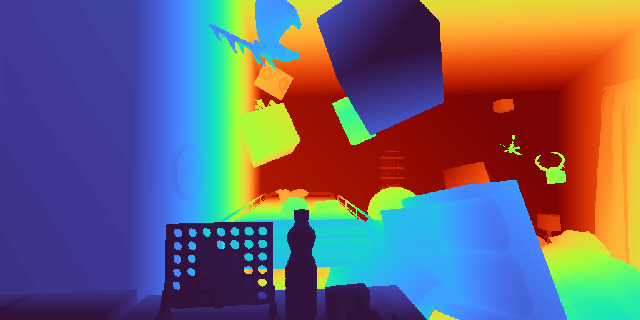}
  \includegraphics[width=\wratio\textwidth]{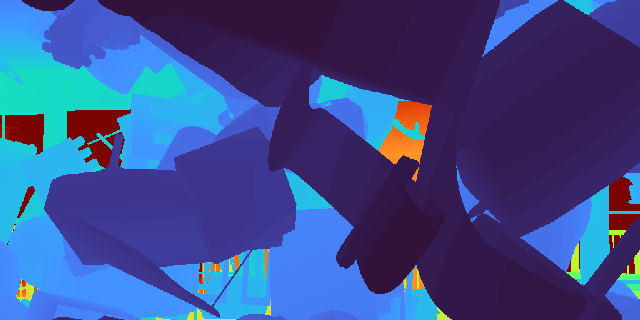}
  \\
  \raisebox{.4cm}{\makebox[0.01\textwidth]{\rotatebox{90}{\normalsize Color}}}
  \includegraphics[width=\wratio\textwidth]{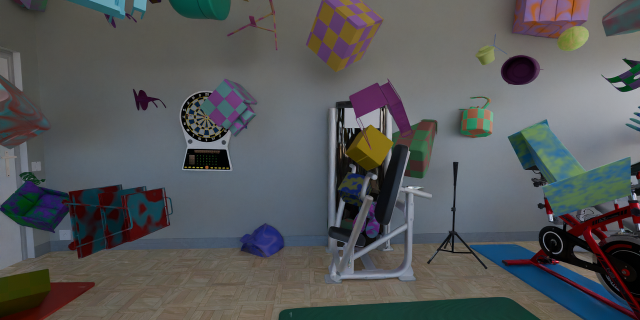}
  \includegraphics[width=\wratio\textwidth]{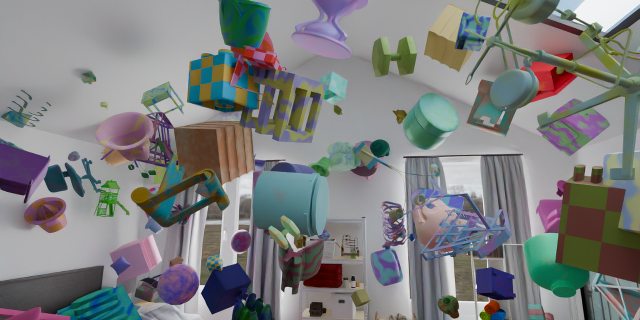}
  \includegraphics[width=\wratio\textwidth]{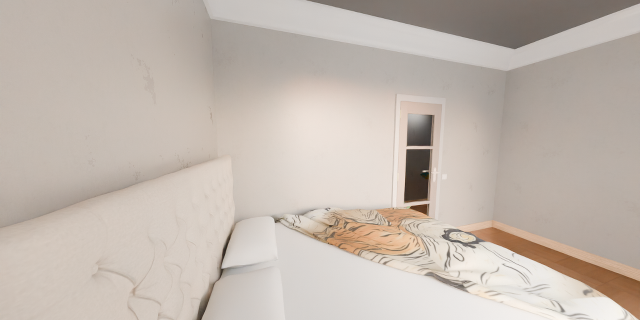}
  \\
  \raisebox{.4cm}{\makebox[0.01\textwidth]{\rotatebox{90}{\normalsize Depth}}}
  \includegraphics[width=\wratio\textwidth]{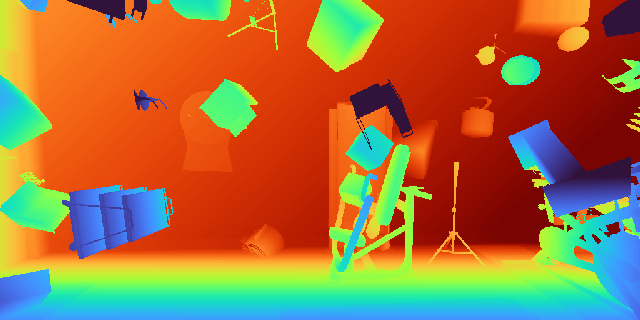}
  \includegraphics[width=\wratio\textwidth]{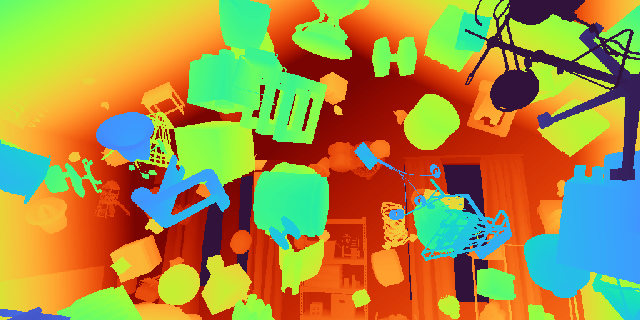}
  \includegraphics[width=\wratio\textwidth]{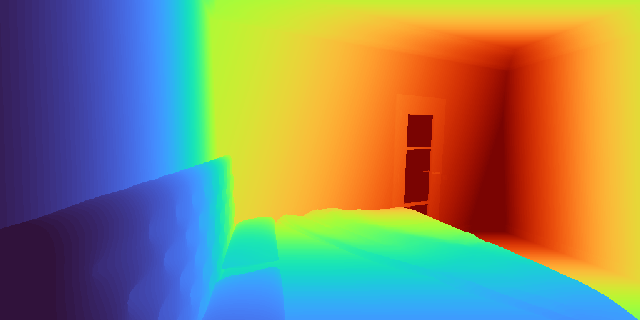}
  \\
  \raisebox{.4cm}{\makebox[0.01\textwidth]{\rotatebox{90}{\normalsize Color}}}
  \includegraphics[width=\wratio\textwidth]{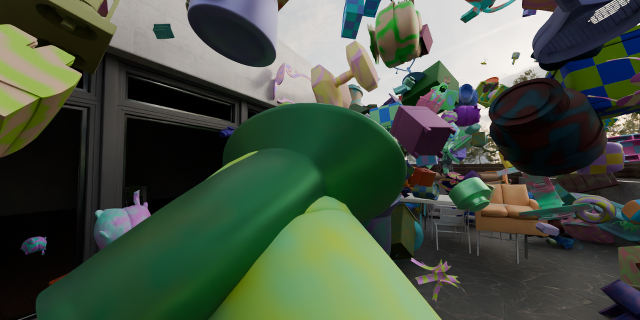}
  \includegraphics[width=\wratio\textwidth]{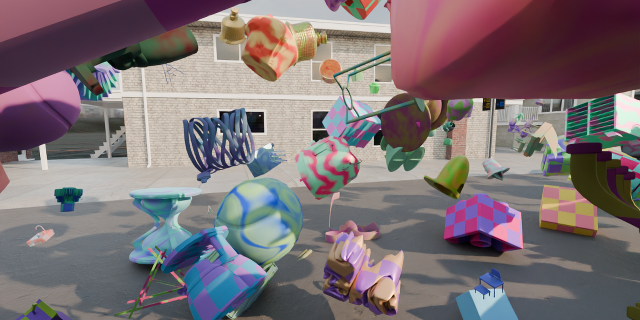}
  \includegraphics[width=\wratio\textwidth]{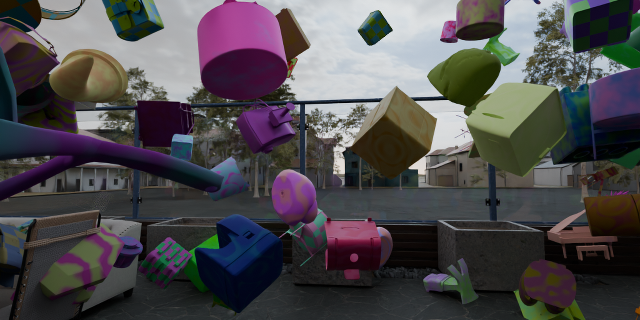}
  \\
  \raisebox{.4cm}{\makebox[0.01\textwidth]{\rotatebox{90}{\normalsize Depth}}}
  \includegraphics[width=\wratio\textwidth]{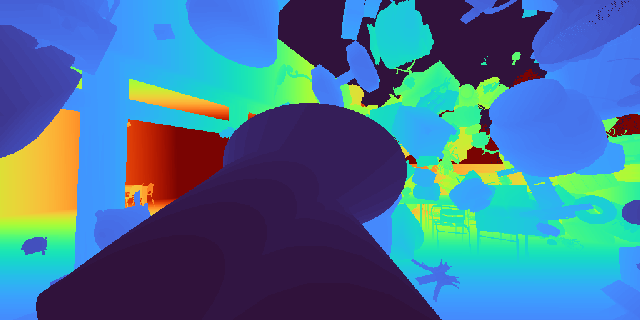}
  \includegraphics[width=\wratio\textwidth]{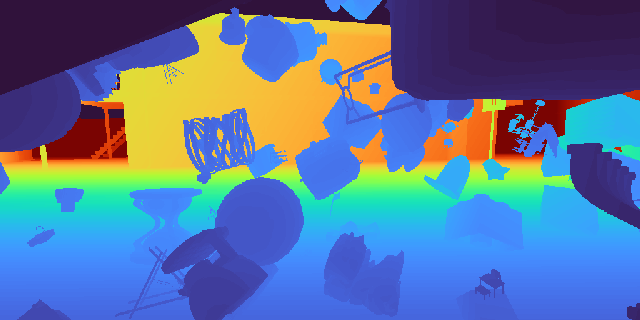}
  \includegraphics[width=\wratio\textwidth]{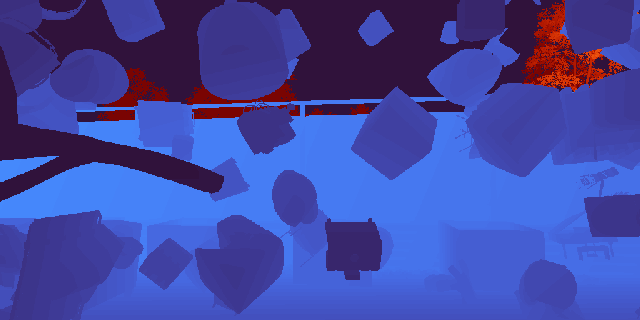}
  \\  
  \caption{\textbf{Samples of \oursdataset}. Representative left-view color frames and corresponding depth maps from our pretraining dataset. \oursdataset comprises 137k high-resolution ($1$K) frames spanning 20k temporal sequences across 13 diverse indoor and outdoor environments populated with 200 unique assets yielding the geometric and photometric diversity needed to train a generalizable feed-forward stereo reconstruction model.}
  \label{supmat_fig:xrstereo_qualitative_results}
\end{figure*}

We provide representative visual samples of \oursdataset in Figure~\ref{supmat_fig:xrstereo_qualitative_results}. To facilitate robust generalization for feed-forward stereo reconstruction, \oursdataset comprises 137k high-quality $1$K resolution frames across 20k temporal sequences. These sequences depict static scenes with dynamic camera trajectories across 13 diverse indoor and outdoor environments, populated with a large-scale library of 200 unique assets.  A distinguishing feature of our dataset is the systematic variation of camera rig configurations, including fields of view ranging from $50^\circ$ to $130^\circ$ and stereo baselines from $40$ to $120\,$mm. This variability ensures the model is resilient to the diverse hardware specifications found in consumer VR/MR headsets and robotic platforms. Furthermore, all samples have passed a multi-stage quality assurance pipeline that filters for sufficient illumination, minimal near-plane clipping, and low occlusion rates. By providing dense, high-fidelity disparity maps and precise camera metadata, \oursdataset offers the geometric complexity necessary for training a robust model.

\FloatBarrier
\subsection{Runtime and Memory Analysis}
\label{supmat_sec:runtime}
We profile all methods on the Infinigen test split under the four-view protocol of Table~\ref{tab:nvs_comparison}, using CUDA-event timing (10 warm-up and 50 timed iterations, batch size 1) on a single NVIDIA A100-80GB GPU. As shown in Table~\ref{supmat_tab:efficiency_comparison}, the end-to-end reconstruction of \ours is currently slower than that of the baselines. However, our aggregation compresses 819k primitives to 132k ($6.2\times$), yielding the lowest peak memory and the highest rendering speed. Since reconstruction runs once per scene while rendering repeats for every target view, rendering throughput is the more relevant metric for interactive novel view synthesis. The efficiency of \ours therefore lies in its compact representation and fast rendering rather than in end-to-end reconstruction latency.

Table~\ref{supmat_tab:runtime_breakdown} shows that this latency is almost entirely due to the frozen FoundationStereo~\citep{wen2025stereo} backbone. Our grounding module, decoder, and aggregation together add only 48.3\,ms, and the aggregation alone, with 34 parameters, adds 4.4\,ms. The backbone cost stems from its iterative GRU refinement and can be tuned without retraining: reducing the iterations from 12 to 8 yields a $1.14\times$ speedup at a $0.6\%$ relative increase in AbsRel, and 6 iterations yield $1.23\times$ at $2.0\%$. Finally, per-pixel primitive generation is linear in the number of pixels and processed per view, and the hash-based aggregation is linear in the number of primitives, so long inputs can be processed in temporal chunks (Fig.~\ref{fig:teaser}).

\begin{table}[h]
\centering
\small
\caption{\textbf{Efficiency comparison} on Infinigen with four input views, measured on a single NVIDIA A100-80GB GPU. Our aggregation yields the most compact representation, the lowest peak memory, and the fastest rendering.}
\label{supmat_tab:efficiency_comparison}
\setlength{\tabcolsep}{5pt}
\begin{tabular}{lcccccc}
\toprule
\multirow{2}{*}{Method} & \multirow{2}{*}{Params} & \multirow{2}{*}{Peak Mem. [GB]$\downarrow$} & \multicolumn{2}{c}{\#Splats} & \multicolumn{2}{c}{FPS$\uparrow$} \\
\cmidrule(lr){4-5} \cmidrule(lr){6-7}
& & & Pre-Agg. & Post-Agg.$\downarrow$ & Total & Render \\
\midrule
DepthSplat~\citep{xu2024depthsplat} & 38.4M & 6.14 & 1,376k & 1,376k & \textbf{5.2} & 59.5 \\
AnySplat~\citep{jiang2025anysplat} & 1.19B & 5.73 & 776k & 505k & 1.9 & 157.2 \\
Ours & 65.0M & \textbf{2.03} & 819k & \textbf{132k} & 1.7 & \textbf{236.9} \\
\bottomrule
\end{tabular}
\end{table}

\begin{table}[h]
\centering
\small
\caption{\textbf{Runtime breakdown} of \ours per reconstruction under the setting of Table~\ref{supmat_tab:efficiency_comparison}. The frozen stereo backbone dominates the runtime, while our modules add only 48.3\,ms.}
\label{supmat_tab:runtime_breakdown}
\begin{tabular}{lccc}
\toprule
Stage & Time [ms] & Share [\%] & Params \\
\midrule
FoundationStereo backbone (frozen) & 558.5 & 91.4 & 62.3M \\
Grounding module + Gaussian decoder & 43.9 & 7.2 & 2.71M \\
Gaussian aggregation & 4.4 & 0.7 & 34 \\
Rendering (post-agg.) & 4.2 & 0.7 & 0 \\
\midrule
Total & 610.9 & 100 & 65.0M \\
\bottomrule
\end{tabular}
\end{table}

\FloatBarrier
\subsection{Alternative Stereo Backbone}
\label{supmat_sec:backbone}
Our framework is not tied to a specific stereo backbone. To demonstrate this, we replace FoundationStereo~\citep{wen2025stereo} with the lighter IGEV++~\citep{xu2023iterative} and keep the rest of the architecture unchanged. As with FoundationStereo, we keep IGEV++ frozen with its publicly released SceneFlow weights and train only the remaining modules. As shown in Table~\ref{supmat_tab:backbone}, the IGEV++ variant still recovers considerably more accurate geometry than all baselines in Table~\ref{tab:nvs_comparison}. It trails our FoundationStereo model by 0.9--3.2\,dB PSNR, with the largest gap on real-world KITTI, showing that a stronger stereo prior directly translates into higher rendering quality.

\begin{table}[h]
\centering
\small
\caption{\textbf{Alternative stereo backbone.} \ours with the FoundationStereo backbone (numbers from Table~\ref{tab:nvs_comparison}) compared to \ours with the IGEV++ backbone, evaluated under the same protocol. Both backbones are frozen.}
\label{supmat_tab:backbone}
\setlength{\tabcolsep}{8pt}
\begin{tabular}{l|ccc|ccc}
\toprule
\multicolumn{1}{c|}{\multirow{2}{*}{Backbone}} & \multicolumn{3}{c|}{Novel View Geometry} & \multicolumn{3}{c}{Novel View Appearance} \\
\cmidrule(lr){2-4} \cmidrule(lr){5-7}
& AbsRel$\downarrow$ & $\delta_1$$\uparrow$ & L1$\downarrow$ & PSNR$\uparrow$ & SSIM$\uparrow$ & LPIPS$\downarrow$ \\
\midrule
\multicolumn{7}{l}{\textit{Infinigen}} \\
IGEV++ & 0.07 & 90.9 & 2.2 & 27.1 & 0.85 & 0.29 \\
FoundationStereo & 0.03 & 98.6 & 1.1 & 29.4 & 0.95 & 0.12 \\
\midrule
\multicolumn{7}{l}{\textit{TartanV2}} \\
IGEV++ & 0.08 & 92.6 & 1.0 & 25.4 & 0.93 & 0.15 \\
FoundationStereo & 0.06 & 94.9 & 0.7 & 26.3 & 0.94 & 0.10 \\
\midrule
\multicolumn{7}{l}{\textit{KITTI}} \\
IGEV++ & 0.10 & 86.5 & 1.8 & 22.0 & 0.88 & 0.22 \\
FoundationStereo & 0.04 & 96.0 & 0.9 & 25.2 & 0.94 & 0.10 \\
\bottomrule
\end{tabular}
\end{table}

\FloatBarrier
\subsection{Gated Depth Refinement}
\label{supmat_sec:gate}
To validate the gated depth refinement, we train variants with the gate $\alpha$ in Eq.~(\ref{eq:depth_refine}) fixed. Setting $\alpha=0$ removes the refinement and renders from the stereo depth $D$ alone, while still using bidirectional inference, the decoder, and aggregation. Setting $\alpha=1$ discards the stereo anchor and renders from the refined depth $\hat{D}$ alone. As shown in Table~\ref{supmat_tab:gate}, the $\alpha=0$ variant retains most of the geometric accuracy but loses 1.3\,dB PSNR and roughly doubles LPIPS. Calibrated stereo thus provides a strong geometric anchor, while the learned refinement mainly improves rendering quality. Conversely, $\alpha=1$ degrades geometry, and neither fixed setting matches the learned gate.

\begin{table}[h]
\centering
\small
\caption{\textbf{Gated depth refinement} on KITTI. Fixing the gate $\alpha$ in Eq.~(\ref{eq:depth_refine}) to either extreme degrades performance compared to the learned gate.}
\label{supmat_tab:gate}
\begin{tabular}{lcccc}
\toprule
Variant & AbsRel$\downarrow$ & $\delta_1$$\uparrow$ & PSNR$\uparrow$ & LPIPS$\downarrow$ \\
\midrule
$\alpha=0$ (no refinement) & 0.042 & 95.8 & 23.9 & 0.20 \\
$\alpha=1$ (no stereo anchor) & 0.050 & 95.7 & 24.8 & 0.14 \\
Ours (learned $\alpha$) & \textbf{0.040} & \textbf{96.0} & \textbf{25.2} & \textbf{0.10} \\
\bottomrule
\end{tabular}
\end{table}

\end{document}